\documentclass[
reprint,
superscriptaddress,
showkeys,
amsmath,
amssymb,
aps,
pre,
floatfix,
hidelinks,
]{revtex4-2}

\usepackage{url}
\usepackage{graphicx}
\usepackage{color}
\usepackage{dcolumn}
\usepackage{bm}
\usepackage{hyperref}

\begin{document}

\title{Breakthrough-Induced Loop Formation in Evolving Transport Networks}

\author{Stanis{\l}aw \.{Z}ukowski} 
    \email{s.zukowski@uw.edu.pl}
    \affiliation{Institute of Theoretical Physics, Faculty of Physics, University of Warsaw, Pasteura 5, 02-093 Warsaw, Poland}
    \affiliation{Laboratoire Mati\`{e}re et Syst\`{e}mes Complexes (MSC), UMR 7057, CNRS \& Universit\'{e} Paris Cit\'{e}, 10 rue Alice Domon et L\'{e}onie Duquet, 75013 Paris, France}
\author{Annemiek Johanna Maria Cornelissen} 
    \affiliation{Laboratoire Mati\`{e}re et Syst\`{e}mes Complexes (MSC), UMR 7057, CNRS \& Universit\'{e} Paris Cit\'{e}, 10 rue Alice Domon et L\'{e}onie Duquet, 75013 Paris, France}
\author{Florian Osselin} 
    \affiliation{Institut des Sciences de la Terre d’Orl\'{e}ans, UMR 7327, CNRS \& BRGM \& Universit\'{e} d’Orl\'{e}ans,  rue de la Ferollerie, 45100 Orl\'{e}ans, France} 
\author{St\'{e}phane Douady}
    \affiliation{Laboratoire Mati\`{e}re et Syst\`{e}mes Complexes (MSC), UMR 7057, CNRS \& Universit\'{e} Paris Cit\'{e}, 10 rue Alice Domon et L\'{e}onie Duquet, 75013 Paris, France}
\author{Piotr Szymczak} 
    \affiliation{Institute of Theoretical Physics, Faculty of Physics, University of Warsaw, Pasteura 5, 02-093 Warsaw, Poland}

\date{\today}

\begin{abstract}
Transport networks, such as vasculature or river networks, provide key functions in organisms and the environment. They usually contain loops whose significance for the stability and robustness of the network is well documented. However, the dynamics of their formation is usually not considered. Such structures often grow in response to the gradient of an external field. During evolution, extending branches compete for the available flux of the field, which leads to effective repulsion between them and screening of the shorter ones. Yet, in remarkably diverse processes, from unstable fluid flows to the canal system of jellyfish, loops suddenly form near the breakthrough when the longest branch reaches the boundary of the system. We provide a physical explanation for this universal behavior. Using a 1D model, we explain that the appearance of effective attractive forces results from the field drop inside the leading finger as it approaches the outlet. Furthermore, we numerically study the interactions between two fingers, including screening in the system and its disappearance near the breakthrough. Finally, we perform simulations of the temporal evolution of the fingers to show how revival and attraction to the longest finger leads to dynamic loop formation. We compare the simulations to the experiments and find that the dynamics of the shorter finger are well reproduced. Our results demonstrate that reconnection is a prevalent phenomenon in systems driven by diffusive fluxes, occurring both when the ratio of the mobility inside the growing structure to the mobility outside is low and near the breakthrough.
\end{abstract}

\maketitle

\section{Significance statement}
Loops are ubiquitous in animate and inanimate transport networks, from leaf venation to river deltas. Yet, current physical models fail to reproduce them accurately. We report a way of loop formation that manifests in remarkably diverse processes -- from unstable fluid flows to the canal system of jellyfish. Near the breakthrough, when one of the branches reaches the outlet of the system, shorter branches revive and accelerate towards the longest one, ultimately leading to a reconnection event. This results in a hierarchical structure of interconnected fingers. Our explanation of the shared underlying mechanism behind the phenomenon significantly advances the understanding of how looping transport networks dynamically emerge.

\section{Introduction}

\begin{figure*}[ht]
\centering
\includegraphics[width=17.8cm]{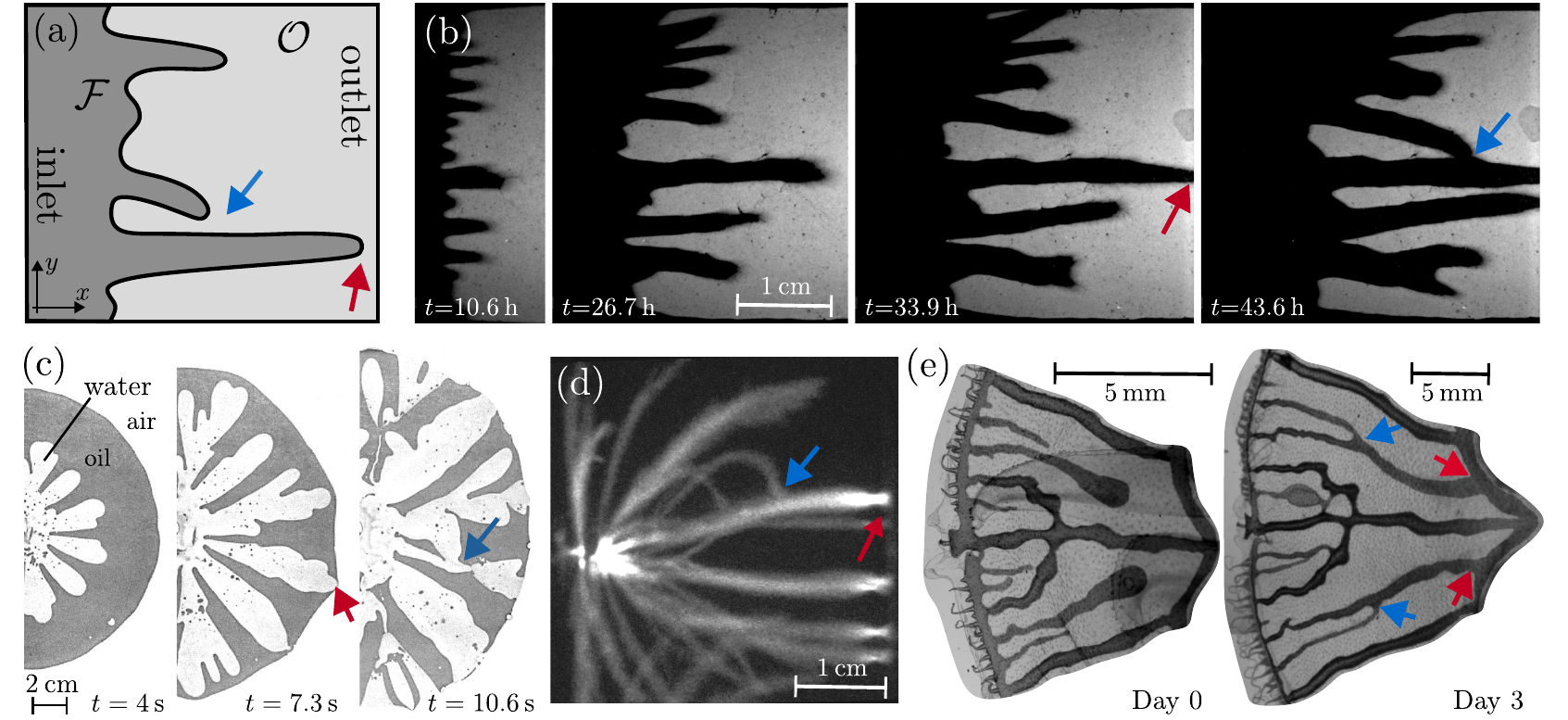}
\caption{ Systems near breakthrough in nature. (a) Each system consists of two phases: an invading phase ($\mathcal{F}$) with mobility $\lambda_1$, and a displaced phase ($\mathcal{O}$) with mobility $\lambda_2$, separated by an interface. The growth direction is from left to right. Red arrows mark places where the fingers are near breakthrough or have already broken through, and blue arrows mark reconnections. (b-e) Examples of reconnection near breakthrough in various systems: (b) a fracture dissolution experiment in a Hele-Shaw cell, (c) viscous fingers in the Saffman-Taylor experiment, (d) streamer channels in air (photo: Sander Nijdam, Eindhoven University of Technology, by permission), (e) an octant of the jellyfish \textit{Aurelia}, showing the gastrovascular canal network in dark gray \cite{song2023morphogenesis}.}
\label{examples}
\end{figure*}

Nature offers us a wide spectrum of spatial, transport networks, which provide key functions in living organisms and the surrounding environment. Examples span from leaf venation \cite{mitchison1980model}, blood vessels \cite{nguyen2006dynamics}, jellyfish gastrovascular canal system \cite{song2023morphogenesis}, to river networks \cite{petroff2011geometry,devauchelle2012ramification}, deltas \cite{konkol2022interplay,ke2019distributary}, or cave conduits \cite{szymczak2011initial,ghani2013dynamic}. Topologically, these networks can take the form of either branched, tree-like structures or looping patterns. The latter are more robust against damage~\cite{katifori2010damage}, hence they are often favored by biological evolution. For example, it is thought that leaf venation is branched in ancestral plants, but highly reticulated in more evolutionary recent ones~\cite{boyce2002evolution,boyce2005evolutionary,douady2020work}.

While the question of why looping networks might have been chosen by evolution and what quantity, if any, they optimize, was extensively studied~\cite{banavar2000topology, bohn2007structure, durand2007structure, katifori2010damage, ronellenfitsch2016global, kaiser2020discontinuous, konkol2022interplay}, the question of how branches in the network interact to form a looping network remains unclear. A natural candidate for an underlying mechanism leading to reconnections is a tensorial stress field~\cite{couder2002leaf,bohn2005hierarchical}. However, many networks in nature develop as a result of unstable growth processes in a scalar field~\cite{fleury2001branching,ball2009branches}.

The growth in such a case is driven by the gradient of an external field, such as electric potential or concentration. Growing parts of the structure compete for the available flux of the field, effectively interacting with each other. The branches in the network try to go away from each other, to maximize the flux coming to their tips and avoid being shadowed~\cite{meakin1998fractals, fleury2001branching, ball2009branches}. This mechanism, however, cannot explain the formation of loops, for which an effective attraction between the growing branches have to be present. 

Surprisingly, loops suddenly appear across numerous systems as the longest branch reaches the outlet of the system. Here we report that near the \textit{breakthrough}, the shorter branch grows toward the longest one and reconnects to it (Fig.~\ref{examples}a). This results in a characteristic hierarchical pattern found in a large variety of systems (Fig.~\ref{examples}b-e): dissolving fractures~\cite{detwiler2003experimental,starchenko2016three}, viscous fingering~\cite{yang2022hydrodynamics}, discharge patterns~\cite{nijdam2009reconnection}, and even the growth of a gastrovascular canal network of the jellyfish \textit{Aurelia}~\cite{song2023morphogenesis}; see \textit{Materials and Methods}, section~\ref{matmethods:exps} for a detailed description of the examples and experiments. The ubiquity of this process suggests the existence of a shared underlying mechanism, which we elucidate here.

\subsection{Laplacian growth} 
In many unstable growth processes, the boundary -- the interface between two phases -- moves due to the external forcing, such as pressure gradient between the inlet and outlet of the system (Fig.~\ref{examples}a). Important parameters in such models are the mobilities of the two phases, e.g. hydraulic permeabilities for pressure driven growth. Whenever the mobility of the invading phase ($\lambda_1$) is larger than the mobility of the displaced phase ($\lambda_2$) the flux concentrates on small protrusions of the interface and the front can break into fingers~\cite{muskat1937flow,ball2009branches,stevens1974patterns}. Because of the flux concentration the fingers tend to grow more in length than in diameter. The width of the fingers is then set by surface tension, or other short-scale regularization mechanisms. Additional effects, such as tip splitting, can give rise to a highly ramified, hierarchical tree-like structure.

A paradigm for such growth processes is Laplacian growth, where the fingers extend with velocity proportional to the flux of a diffusive field ($\phi_i$) given by: $\vec{J_i}=-\lambda_i \nabla \phi_i$, for $i=1,2$ depending on the phase. The conservation of the flux results in the Laplace equation for the field in both the invading ($\mathcal{F}$) and the displaced phase ($\mathcal{O}$):
\begin{equation}
    \begin{aligned} \label{eq1}
    \Delta \phi_1(\vec{x}) = 0 \ \ \vec{x}\in \mathcal{F} \ \ \ \ \textrm{and} \ \ \ \
    \Delta \phi_2(\vec{x}) = 0 \ \ \vec{x}\in \mathcal{O} \,.
    \end{aligned}
\end{equation}
These equations are supplemented with the continuity condition for the field and its flux at the interface ($\mathcal{T}$):
\begin{equation}\label{laplace2}
    \begin{aligned}
    \phi_1(\vec{x}) = \phi_2(\vec{x}) \ \ \ \ \ \ \ &\vec{x} \in \mathcal{T} \,, \\
    \lambda_1 (\nabla \phi_1(\vec{x}))_\textrm{n} = \lambda_2 (\nabla \phi_2(\vec{x}))_\textrm{n} \ \ \ \ \ &\vec{x} \in \mathcal{T} \,,
    \end{aligned}
\end{equation}
where $n$ denotes the normal to the interface. Additionally, the Dirichlet boundary condition is imposed on the inlet ($x=0$) and outlet ($x=1$) of the system:
\begin{equation}
    \begin{aligned}\label{eq_last}
    \phi_1(x=0) = 1 \ \ \  \textrm{and} \ \ \ 
    \phi_2(x=1) = 0 \,.
    \end{aligned}
\end{equation}
Note that the coordinates here are rescaled by the system length, and the field is rescaled by the value at the inlet.

\subsection{Effect of mobility ratio}
However, reticulated networks are usually not obtained in such models~\cite{meakin1998fractals, fleury2001branching, ball2009branches}. This is due to the simplifying assumption that neglects the mobility of the invading phase, which is equivalent to taking the limit of mobility ratio going to infinity, $M=\lambda_1 / \lambda_2 \to \infty $. This leads to the omission of the field drop within the fingers. The Laplace equation is then solved only in the displaced domain with a constant value of the field directly on the moving boundary. The Dirichlet boundary condition on the fingers results in: (i) long fingers screening the shorter ones and hindering their growth; (ii) two parallel fingers growing away from each other as they get more flux from the sides.

If one takes into account a finite mobility ratio the effective repulsion and screening diminishes. The field inside the invading phase is then no longer constant and the resulting field gradients can make the fingers attract each other and create loops~\cite{luque2014growing,budek2017effect}. As shown by Budek et al.~\cite{budek2017effect}, this can occur only for a specific range of mobility and finger length ratios. 

However, as presented in Fig.~\ref{examples}, the breakthrough reconnections are not limited to these specific scenarios. Notably, they can occur even in the general case of high mobility ratio, which was previously thought to be impossible due to screening effects. Nonetheless, near the breakthrough, screening diminishes, enabling the revival and growth of shorter branches towards the longest one, ultimately leading to a reconnection event. We demonstrate that contrary to previous studies reconnection is a prevalent phenomenon in Laplacian growth, occurring both in low mobility ratio cases and near the breakthrough.

\section{Results and Discussion}

\subsection{Drop of potential along a single finger}
\begin{figure}[t]
\centering
\includegraphics[width=1\linewidth]{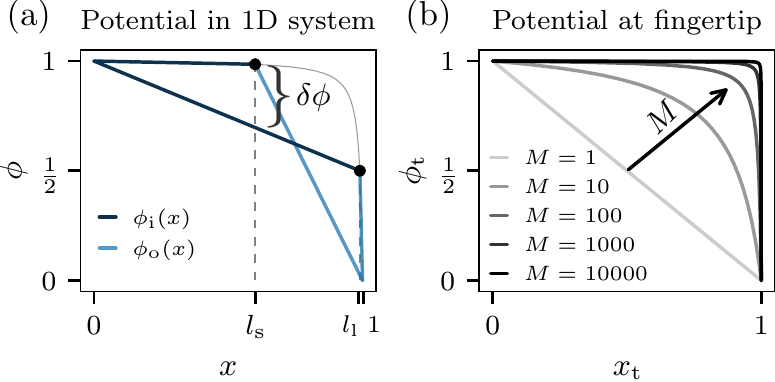}
\caption{Solutions of Eqs.~(\ref{eq1})-(\ref{eq_last}) for a 1D system. (a)~Potential in the system with mobility ratio $M=100$ for a short finger of length $l_\textrm{s}$ and a longer finger of length $l_\textrm{l}>l_\textrm{s}$. The dark line marks the field inside the finger ($\phi_\textrm{i}$), the lighter line marks the field outside ($\phi_\textrm{o}$), and the black dot marks the position of the fingertip. (b)~A profile of the potential at the fingertip ($\phi_\textrm{t}$) as a function of the tip position ($x_\textrm{t}$) for different mobility ratios. Such a profile for $M=100$ is also plotted in (a) with a thin gray line.}
\label{1dsolutions}
\end{figure}

To understand the generality of the breakthrough reconnections we begin with a 1D case. Here, $\mathcal{F}=\{ x \in [ 0, x_\textrm{t}] \}$, $\mathcal{O}=\{ x \in [x_\textrm{t}, 1] \}$ and $\mathcal{T} = \{x_\textrm{t}\}$ is just the fingertip. The solution of the Eqs.~(\ref{eq1})-(\ref{eq_last}) are then two piecewise linear functions stitched at the fingertip (Fig.~\ref{1dsolutions}a)
\begin{equation}
    \phi(x) = 
\begin{cases}
    1 - \frac{x}{M - x_\textrm{t} (M - 1)}, & \text{if } x \in [ 0, x_\textrm{t}]\\
    \frac{1 - x}{1 - x_\textrm{t} (1-1/M)},              & \text{if } x \in [x_\textrm{t}, 1].
\end{cases}
\end{equation}
A few remarks can be made based on this simple example.

First, for a finite mobility ratio ($M=100$, Fig. \ref{1dsolutions}a) when the fingertip is far from the outlet ($x_\textrm{t} = l_\textrm{s} \ll 1$), approximation of a constant field on the moving boundary works well, and the field inside the short finger can be treated as constant. However, if we take a longer finger ($x_\textrm{t} = l_\textrm{l} \approx 1$), there is a significant potential drop inside it. By comparing the field values at $x=l_\textrm{s}$ we see that there is a difference of potential between the two fingers:
\begin{equation}\label{deltaphi}
    \delta\phi \approx 1-\phi(x=l_\textrm{s},x_\textrm{t}=l_\textrm{l})=\frac{l_\textrm{s}}{1 + d (M-1)},
\end{equation}
where $d=1-l_\textrm{l}$. If such fingers were placed at a distance $\delta y$ next to each other, there would be a gradient of potential $\delta \phi / \delta y$, between the shorter finger tip and the longer finger, provided that the fingers do not influence each other. As a result, the shorter finger would be attracted toward the longer one.

Second, let us focus on the potential at the fingertip as a function of the tip position $\phi_\textrm{t}(x_\textrm{t}) = \phi(x=x_\textrm{t})$ (Fig.~\ref{1dsolutions}a, light gray line and Fig.~\ref{1dsolutions}b), and analyze how it changes with the mobility ratio. The higher the $M$, the steeper the profile of the potential at the fingertip, and the longest finger must be closer to the outlet to feel its impact. As can be seen in \eqref{deltaphi}, $\delta\phi$ is inversely proportional to the product $Md$. This suggest that the critical distance,~$d_\textrm{c}$, at which the pressure inside the longer finger begins to drop and the attraction between two fingers would appear, scales as $d_\textrm{c} \sim M^{-1}$.

In the limiting case $M \to \infty$ the functional dependence $\phi_\textrm{t}(x_\textrm{t})$ becomes a step function (Fig.~\ref{1dsolutions}b). The fingers have a constant potential along their length, no matter how close they are to the outlet, and there is no difference in potential between them. However, when one of the fingers breaks through, the potential inside it takes the form $\phi(x) = 1 - x$. This instantaneous pressure drop inside the longer finger induces a pressure difference with respect to the shorter finger, $\delta \phi=l_\textrm{s}$. Consequently, at the moment of breakthrough, we observe a sudden transition from no interaction between the fingers to attraction of the shorter finger to the longer one.

\subsection{Finger interactions in two dimensions---screening and revival}

\begin{figure}[t]
\centering
\includegraphics[width=8.6cm]{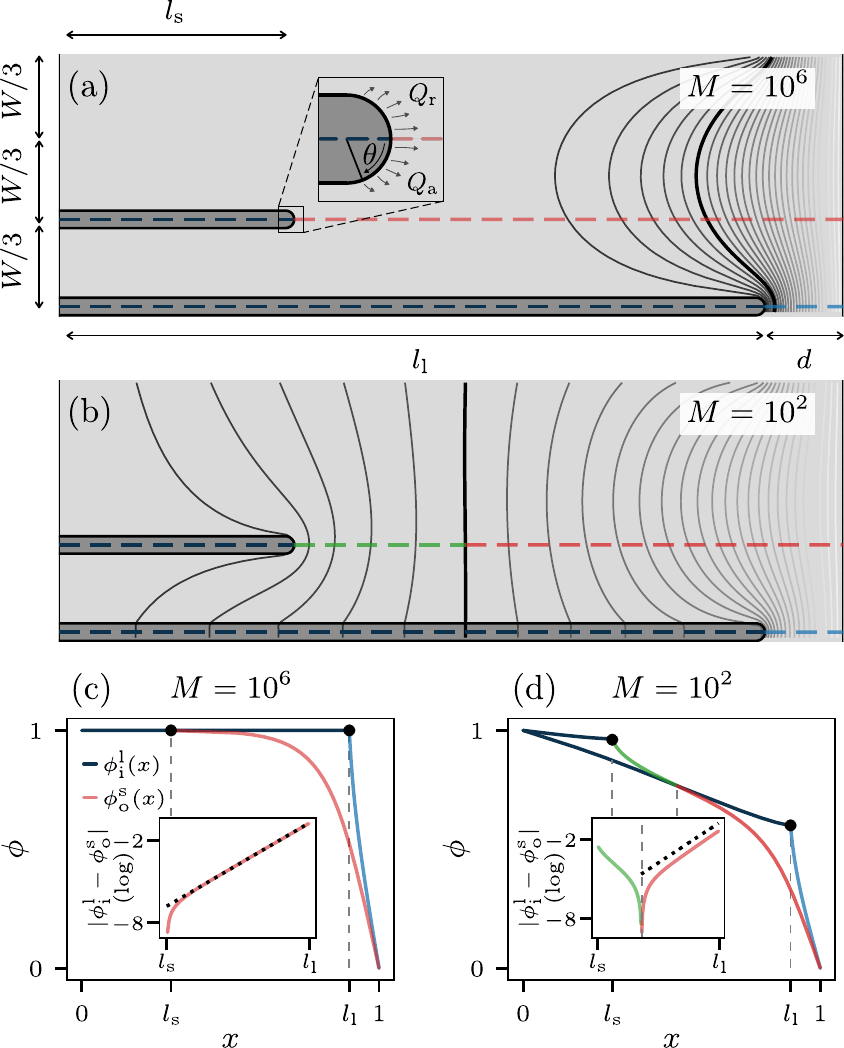}
\caption{ Numerical solutions of Eqs.~(\ref{eq1})-(\ref{eq_last}) for a 2D system. (a)-(b)~A schematic of the 2D setup with two fingers (of lengths $l_\mathrm{s}=0.3$ and $l_\mathrm{l}=0.9$) in a periodic cell of length $L=1$ and width $W=L/3$ with the isolines of the field for: (a)~$M=10^2$ and (b)~$M=10^6$. In the inset we mark the attractive ($Q_\textrm{a}$) and repulsive ($Q_\textrm{r}$) fluxes intercepted by the right and the left side of the fingertip, respectively. \mbox{(c)-(d)}~Field values along the longer and shorter finger (cross sections marked with dashed lines in (a) and (b)) for the two mobility ratios. Insets in (c)-(d): the difference between the field inside the longer finger and outside the shorter one: $\phi_\textrm{i}^\textrm{l}~-~\phi_\textrm{o}^\textrm{s}$. The black dashed line is a fit to the linear part of the plot in inset (c), highlighting the exponential decay of $\phi_\textrm{i}^\textrm{l}~-~\phi_\textrm{o}^\textrm{s}$ (from the longer finger tip perspective).}
\label{2dfingers}
\end{figure}

Although insightful, the 1D model treats two fingers independently of each other, and does not take into account effects of finger interactions such as screening. To investigate these effects we conduct numerical simulations in a 2D geometry. Two fingers are placed in a cell of length $L=1$ and width $W=L/3$ with periodic boundary conditions on the bottom ($y=0$) and top ($y=W$) wall. The long finger of length $l_\textrm{l}$ is positioned at $y=0$, and the short finger of length $l_\textrm{s}$ is at $y=W/3$, as shown in Figs.~\ref{2dfingers}a-b. The fingers have a shape of thin rectangles of width $W/15$ with semicircular caps. We solve the equations for the field with the finite element method implemented in the FreeFEM++ software~\cite{hecht2012new} (\textit{Materials and Methods}, section~\ref{matmethods:numerical}). 

Figs.~\ref{2dfingers}a-b represent isolines of the field (the same set of values) in two cases: $M=10^6$ and $M=10^2$. We observe that for high mobility ratio there is a negligible potential drop inside the fingers, and the longer finger attracts almost all of the flux. As a result, it screens the shorter one and would suppress its growth. For lower mobility ratio, due to the potential drop inside the longer finger, some flux can reach the shorter one, so it can still grow.

We consider the field values along the center line of the long and short fingers for the two mobility ratios (Figs.~\ref{2dfingers}c-d). The profiles along the longer finger are piecewise linear, similar to the 1D solutions presented in Fig.~\ref{1dsolutions}a. However, the field values outside the shorter fingers ($\phi_\textrm{o}^\textrm{s}$) are strongly influenced by the longer ones. In fact, the difference between the field inside the longer finger and the field outside the shorter one, $\phi_\textrm{i}^\textrm{l} - \phi_\textrm{o}^\textrm{s}$, decays exponentially from the perspective of the longer finger tip, as shown by the black dashed lines in the insets in Figs.~\ref{2dfingers}c-d. As a result, the shorter finger is strongly screened and almost no flux reaches its tip.

\begin{figure*}[ht]
\centering
\includegraphics[width=17.8cm]{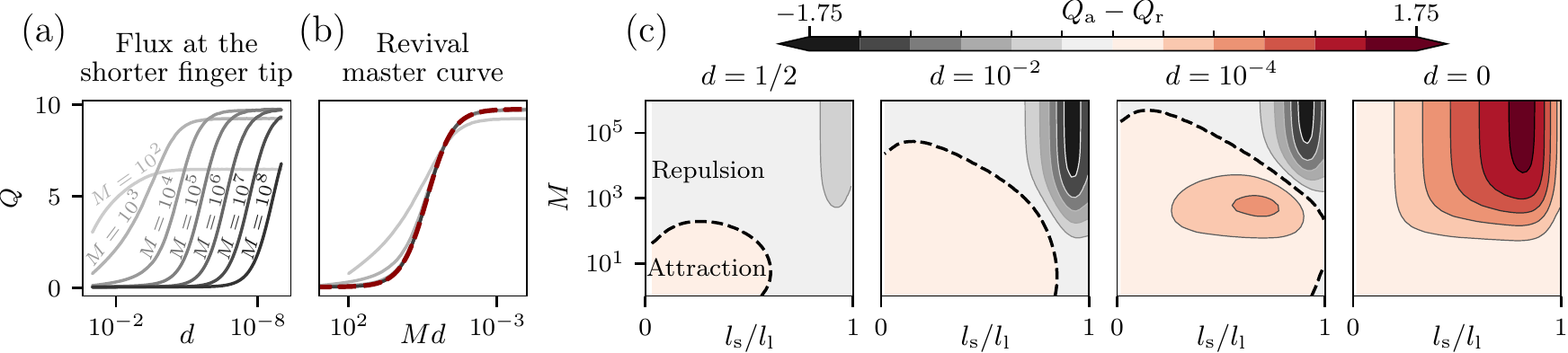}
\caption{Revival and interactions between the fingers when the gap changes. (a) Total flux intercepted by the shorter finger tip, $Q=Q_\textrm{a}+Q_\textrm{r}$ (as defined in inset in Fig.~\ref{2dfingers}a), as a function of $d$. The length ratio is kept constant, $l_\textrm{s}/l_\textrm{l}=1/3$. (b) Total flux plotted as a function of $Md$. The master curve, red dashed line, corresponds to \eqref{master} plotted with $q_0=0.05 , q_1=9.76, d^\prime_\textrm{c}=0.23$. (c)~Maps of interactions between the fingers, $Q_\textrm{a}-Q_\textrm{r}$, as a function of two parameters: mobility ratio ($M$) and finger length ratio ($l_\textrm{s}/l_\textrm{l}$). Maps are plotted for four gaps:  $d= 1/2, \ 10^{-2}, \ 10^{-4}, \ 0$. Red colors indicate attraction, gray repulsion, black dashed line separates the two. }
\label{repulsion_attraction_map}
\end{figure*}

The situation changes drastically as the longer finger approaches the outlet. In Fig.~\ref{repulsion_attraction_map}a we analyze the total flux $Q$ through the tip of the shorter finger as a function of gap $d$ -- the distance from the longer finger tip to the outlet of the system. The length ratio is kept constant $l_\textrm{s}/l_\textrm{l}=1/3$ and $Q$ is calculated by integrating the flux at the shorter finger tip for $\theta \in [-\pi/2, \pi/2]$, see inset in Fig.~\ref{2dfingers}a. 

For relatively low mobility ratio, the shorter finger is weakly screened and receives some flux, even when the longer finger is far from the outlet (large $d$). The higher the $M$, the stronger the screening effect, hence for large $d$ the total flux in the shorter finger is close to zero, and it only starts to increase as the gap gets smaller. Note that the critical distance from the outlet, at which the screening starts to diminish, decreases as $M$ gets larger. This is a further manifestation of the fact that the critical gap at which screening vanishes scales as $M^{-1}$, as already described with the 1D model. The above confirms that for sufficiently small $d$, or after the breakthrough, short fingers that were strongly screened would revive and start growing again, as also noted in Ref.~\cite{starchenko2016three,yang2022hydrodynamics}.

Interestingly, for large $M$ the $Q(d)$ dependence approaches a universal function that simply shifts in $d$ as $M$ increases. This suggests that the flow profiles should collapse onto a master curve when plotted as a function of $Md$, which is indeed the case (Fig.~\ref{repulsion_attraction_map}b). Moreover, if we define rescaled $d$: $d^\prime=Md$, plug it into \eqref{deltaphi}, and take the limit of $M \to \infty$ we see that the master curve will be of the form $1/(1+d^\prime)$. This can be adapted for a 2D system and written as 
\begin{equation}\label{master}
    Q(d^\prime)=q_0+(q_1-q_0)/(1+d^\prime/d^\prime_\textrm{c}).
\end{equation} 
The three parameters $q_0, q_1, d^\prime_\textrm{c}$ can be easily extracted from the simulation results and have a clear physical interpretation: $q_0$ and $q_1$ are the flux reaching the shorter finger tip during screening phase and after revival, respectively; $d^\prime_\textrm{c}$, the inflection point of the sigmoid, is the rescaled critical gap at which the shorter finger in the two-dimensional system revives. As can be seen in Fig.~\ref{repulsion_attraction_map}b, dashed line, \eqref{master} perfectly captures revival in our system.

\subsection{Finger interactions in two dimensions---repulsion and attraction}
Having described the revival of the shorter finger, let us discuss the change in their growth direction near the breakthrough. To quantify this, we calculate the attractive, $Q_\textrm{a}$, and repulsive flux, $Q_\textrm{r}$, by integrating the flux on the shorter finger tip for $\theta \in [0, \pi/2]$ and $\theta \in [-\pi/2, 0]$, respectively (see inset in Fig.~\ref{2dfingers}a). We then take the difference of the two fluxes, $Q_\textrm{a} - Q_\textrm{r}$. The positive value of this quantity will be interpreted as an effective attraction, and should result in the shorter finger growing toward the longer one. Conversely, if the value is negative the fingers repel. Fig.~\ref{repulsion_attraction_map}c represents repulsion-attraction maps as a function of two parameters: mobility ratio ($M$) and finger length ratio ($l_\textrm{s}/l_\textrm{l}$). The maps are shown for four gaps: $d= 1/2, \ 10^{-2}, \ 10^{-4}, \ 0$; the latter corresponds to the breakthrough. 

For large gap (Fig.~\ref{repulsion_attraction_map}c, $d=1/2$), we observe an island of attraction in the region of lower mobility ratio, $M \in (10^0-10^2)$, and the length ratio, $l_\textrm{s}/l_\textrm{l} \in (0, 0.6)$. This quantitatively agrees with the results of the resistor model presented by Budek et al.~\cite{budek2017effect}, where the interactions between the fingers distant from the outlet were analyzed. As described there and as can be seen in Fig.~\ref{repulsion_attraction_map}c, $d=1/2$, for the systems of low mobility ratio the screening is weak and the fingers can attract each other to form loops. Note that in the other maps in Fig.~\ref{repulsion_attraction_map}c, regardless of the gap size $d$, the value of $Q_\textrm{a} - Q_\textrm{r}$ remains almost the same in the region of low mobility and length ratio. This suggests that in such systems the breakthrough will not drastically affect the dynamics and interactions between the fingers.

The impact of the gap size on the interactions is more apparent in the region of higher mobility ratio, $M \in (10^3-10^6)$, and length ratio, $l_\textrm{s}/l_\textrm{l} \in (0.6, 0.9)$. Here, the interactions change from slight to strong repulsion when the leading finger is distant from the outlet, $d=1/2, \ 10^{-2}$, and finally, after the breakthrough, $d=0$, we observe a transition from strong repulsion to strong attraction. Hence, for systems of higher mobility ratio the impact of the breakthrough should be more striking in the dynamics of the fingers. In particular, even for very high mobility ratio, the longer finger will always attract the shorter ones after the breakthrough.

\subsection{Temporal evolution of the fingers}
Finally, we perform dynamic simulations of finger growth. During growth, we do not change the finger shape (constant width and semicircular tips) and extend it only in the direction from which the highest flux is coming (\textit{Materials and Methods}, section~\ref{matmethods:numerical}). Here we analyze the evolution of two fingers with slightly different initial lengths, $l_\textrm{s}=0.04$ vs $l_\textrm{l}=0.05$, and mobility ratio in the system $M=1000$. With such initial conditions we can observe all previously described interactions between the fingers: competition and repulsion, screening, revival and attraction. The dynamics of the fingers are presented in a supplementary video \textit{Movie S1}.

\begin{figure}
\centering
\includegraphics[width=1\linewidth]{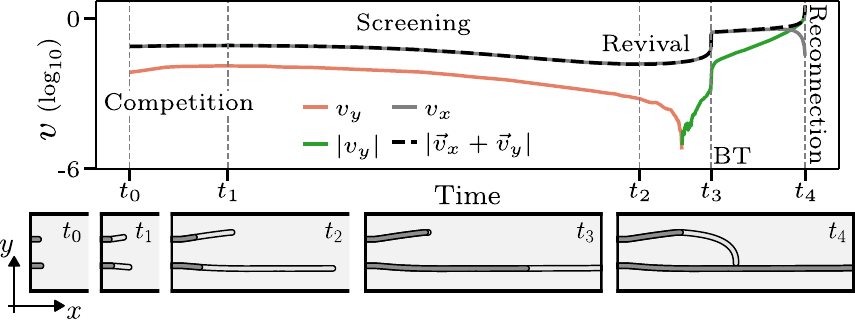} 
\caption{Velocity of the shorter finger in the dynamical simulation of growth with mobility ratio $M=1000$. Black dashed line marks the total velocity of the finger, gray its $x$-component (pointing towards the outlet), red the $y$-component (repulsion from the longer finger) and green the absolute value of $v_y$ (attraction to the longer finger). Gray vertical lines mark: initial time ($t_0$); start of screening ($t_1$), when the acceleration of the finger becomes negative; start of revival ($t_2$), when the acceleration changes its sign again; breakthrough (BT, $t_3$) and reconnection ($t_4$). Snapshots from the simulation were taken at the corresponding moments of the evolution (lighter color on the fingers marks the parts that grew between successive moments).}
\label{velocity}
\end{figure}

In Fig.~\ref{velocity} we show how the velocity of the shorter finger changes over time, along with the snapshots from the simulation. First, as the fingers start with similar lengths, we observe competition and repulsion between them -- the fingers accelerate and grow with similar rate, until the longer finger wins and starts to screen the shorter one. At this moment ($t=t_1$), the shorter finger starts to slow down. As the longer finger approaches the outlet, the field inside it begins to decrease (as visible in \textit{Movie S1}). Consequently, screening disappears and the overall flux within the shorter finger increases. The shorter finger revives and accelerates again ($t=t_2$).

Shortly after $t_2$ the attraction towards the longer finger appears: the positive $v_y$, away from the longer finger (depicted by the red line in Fig.~\ref{velocity}) transitions to negative, toward the longer finger ($|v_y|$, marked with the green line in Fig.~\ref{velocity}). This stage corresponds to the expansion of the attraction region on the interaction maps prior to the breakthrough (Figs.~\ref{repulsion_attraction_map}a). Then, just before the breakthrough ($t=t_3$), there is a sudden jump in velocity. This last boost significantly expedites the loop formation process. As the distance between the two fingers diminishes, the attraction intensifies. This cumulative, snowball-like effect, triggered by the breakthrough, culminates in the eventual reconnection at $t=t_4$.

\subsection{Comparison of the simulations to experiments}
\begin{figure*}
\centering
\includegraphics[width=17.8cm]{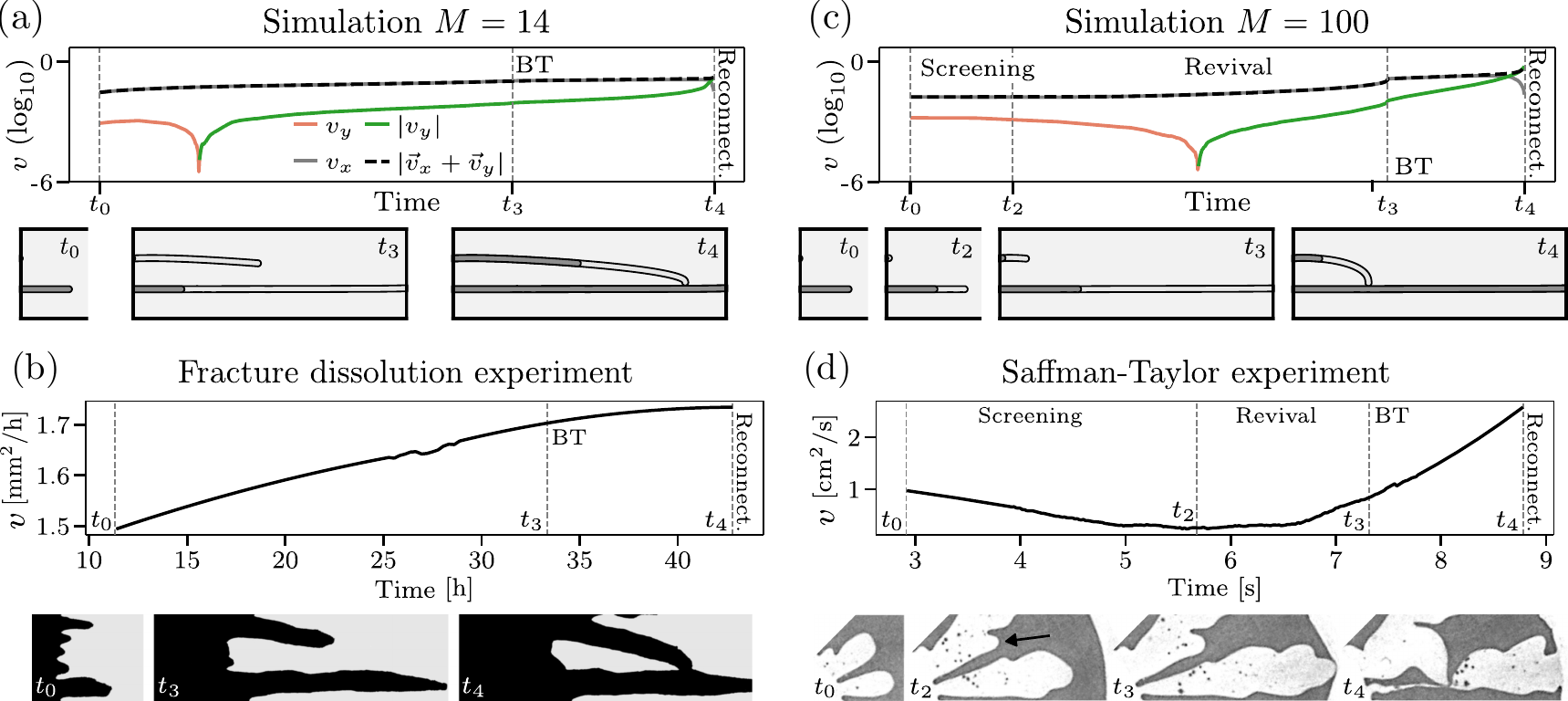}
\caption{ Comparison of simulations to experiments. (a)-(b)~Velocities of the shorter finger in the fracture dissolution simulation and experiment, respectively. The mobilities here are related to the aperture values in the dissolved and undissolved regions (\textit{Materials and Methods}, section~\ref{matmethods:fracture}). The mobility ratio is $M = 14$. (c-d)~Velocity of the shorter finger in the viscous fingering simulation and experiment. Here, the mobility ratio is the inverse of the viscosity ratio and is $M = 100$. The line colors and time points in the plots indicate the same as in Fig.~\ref{velocity}. Snapshots were taken at the corresponding moments of the evolution. The lighter color on the fingers from the simulations marks the parts that grew between successive moments. The black arrow in (d) marks the shorter finger.}
\label{velocity_exps}
\end{figure*}

We additionally present two simulations that qualitatively reproduce the behavior of the fingers in the fracture dissolution and Saffman-Taylor experiments seen in Figs.~\ref{examples}b-c, with approximate mobility ratios of $M = 14$ and $M=100$, respectively (see \textit{Materials and Methods}, section~\ref{matmethods:fracture} and ~\ref{matmethods:viscous} for details on how the mobility ratio in the experiments is calculated). In these simulations, the fingers were initiated with lengths of $l_\textrm{s}=0.002$ and $l_\textrm{l}=0.18$. In \textit{Supplemental Material} we include videos from the simulations (\textit{Movie S2} and \textit{Movie S3}) and videos from the experiments (\textit{Movie S4} and \textit{Movie S5}). In Fig.~\ref{velocity_exps} we show the plots of the velocity of the shorter finger, both in the simulations and in the experiments, in the case of fracture dissolution ($M=14$) and viscous fingering ($M=100$).

As predicted in the previous section, in the case of the lower mobility ratio, the shorter finger is attracted towards the longer one almost from the very beginning of the evolution, both in the simulation (Fig.~\ref{velocity_exps}a, \textit{Movie S2}) and the experiment (Fig.~\ref{velocity_exps}b, \textit{Movie S4}). There is also no effect of screening, and flux is non-negligible even when the longer finger is far away from the outlet. As a result, the finger grows with an almost constant total velocity, which remains unchanged even during the breakthrough.

In contrast, for higher mobility ratio, the shorter finger in the simulation (Fig.~\ref{velocity_exps}c, \textit{Movie S3}) and in the experiment (Fig.~\ref{velocity_exps}d, \textit{Movie S5}) is initially screened by the longer one and grows relatively slowly. Only after the breakthrough does it revive and eventually reconnect forming a loop.

\section{Conclusions}
As previously described in the literature, attractive interactions between the fingers can appear in the systems of low mobility ratio, leading to reconnections. For high mobility ratio, only screening and repulsion have been observed, resulting in branched loopless structures. However, when studying systems such as viscous fingering or fracture dissolution, experiments are often terminated when the invading phase reaches the border of the system. An unexpected behavior arises in a broad class of unstable growth processes when the proximity of the outlet is considered and the evolving structure breaks through.

We have shown that a striking transition in the system dynamics occurs when the breakthrough is reached, especially in the case of infinite mobility ratio, where a singular limit emerges. Prior to the breakthrough, the field along the longest finger remains constant regardless of its proximity to the outlet, and the finger screens the rest of the system. At the moment of breakthrough, however, the field in the longer finger drops dramatically, allowing the flux to reach the shorter neighboring fingers. This affects the dynamics of the shorter fingers, causing their revival and a strong attraction to the longer finger. The interplay of these two factors leads to reconnection and loop formation.

To explain breakthrough reconnections, we used a simplified model of growing Laplacian fingers: (i) we neglected the jump in the field across the interface associated with the regularization mechanism; (ii) we assumed that the change of mobility across the interface is discontinuous, rather than smooth as in, for instance, fracture dissolution; (iii) we kept the shape of the fingers unchanged during their evolution, whereas in some systems, such as viscous fingering, when the fingers receive more flux, their tips begin to grow in width, which impacts their velocity~\cite{bischofberger2015island,zik1999fingering}. Regardless, for high mobility ratio, boundary condition on the fingers transitions from a constant field in the initial state to a linear gradient near the breakthrough. This field change near the breakthrough is much larger than any jump in the field resulting from regularization mechanisms (\textit{Materials and Methods}, section~\ref{matmethods:viscous}). The transition between the two boundary conditions might be more gradual due to (ii) and (iii), but it still significantly impacts the finger dynamics. This is evidenced by sudden loop formation observed in physical systems such as viscous fingering or discharge patterns, and also in living organisms such as the jellyfish \textit{Aurelia} (Figs.~\ref{examples}b-e). 

The breakthrough reconnections are expected to occur in virtually any system driven by diffusive fluxes. Observing it in a system of a yet unknown growth mechanism, such as the gastrovascular canal network of the jellyfish (Fig.~\ref{examples}e), is a strong indication that the system dynamics are controlled by the effective diffusion of a morphogenic parameter. This sheds new light on the possibility of dynamical loop formation in many systems.

\section{Materials and Methods}

\subsection{Experiments}\label{matmethods:exps}
\subsubsection{Fracture dissolution}\label{matmethods:fracture}
The fracture dissolution experiment presented in Fig.~\ref{examples}b was performed in a microfluidic setup described in detail in Ref.~\cite{osselin2016microfluidic}. It consists of two polycarbonate disks. The bottom one contains a rectangular indentation (3.3~cm $\times$ 3.8~cm $\times$ 100~$\mu$m), which is initially filled with a soluble material, in our case plaster (Plaster of Paris, Blik Modelarski Alabastrowy). The top plate contains a hierarchical system of inlet and outlet channels connected to large inlet/outlet reservoirs (4.5~cm $\times$ 5~mm $\times$ 2~mm). Such a design helps to maintain uniform pressure across the width of the plaster. The aperture above the plaster is created by gluing the plates together with an ultrathin PET-based double-coated tape 70~$\mu$m thick with a rectangular hole the size of the plaster block. The cast was prepared with a 60\% (w/w) ratio of water to plaster. This yields an average porosity of the block of $\phi = 50\%$ (measured porosity to water) and a permeability of 45~mD (measured by injection of isopropanol). Pure water is injected into the system with a syringe pump (Harvard Apparatus PHD2000) at a rate of $q=0.5\textrm{~ml/h}$. We recorded the experiment with a UI 1550LE-C-HQ CCD camera (IOS, Germany), acquiring photographic images of the system every 100 s. In order to ensure homogeneous light intensity over the system, we used a circular fluorescent illuminator.

For experiments conducted in a Hele-Shaw cell, the mobility can be expressed as $\lambda = h^3/12\mu$, where $h$ is the aperture of the Hele-Shaw cell available to the fluid and $\mu$ is the fluid viscosity. Thus, increasing the aperture in the dissolved part of the system (black area in Fig.~\ref{examples}b) effectively introduces two phases with different mobilities. The mobility ratio in this experiment was approximately $M = (h_1/h_2)^3 = (170/70)^3 \approx 14$.

\subsubsection{Saffman-Taylor experiment} \label{matmethods:viscous}
The Saffman-Taylor experiment presented in Fig.~\ref{examples}c was performed in a circular Hele-Shaw cell with a 1~mm separation between the bottom and top plates. The Hele-Shaw cell was initially filled with oil (dyed with paprika to increase contrast in the images). Water was injected into the system through an inlet located in the center of the top plate. The experiment was recorded with a Nikon D3000 camera. A LED panel was used below the Hele-Shaw cell to ensure uniform light intensity throughout the system. the mobility ratio is the inverse of the viscosity ratio and was of the order of $M = \mu_2 / \mu_1 \approx 100$.

The pressure jump in a Hele-Shaw is estimated using the Young-Laplace equation: $\Delta p^\textrm{cap}=\gamma(2/b+1/r)$, where $\gamma=49$ mN/m is the surface tension between oil and water, $b=1$ mm is the spacing between the plates, $r\sim1$ cm is a typical radius of curvature for the fingers. This yields $\Delta p^\textrm{cap}\sim10^{-3}$ Pa. When the breakthrough occurs, the pressure inside the longest finger drops from the inlet value to a linear gradient between the inlet and outlet. The typical pressure difference between inlet and outlet in our Saffman-Taylor experiment was of the order of $10^2$ Pa. Thus, the change in $\phi$ induced by the breakthrough is 5 orders of magnitude larger than the pressure jump due to surface tension, and it can be neglected.

\subsubsection{Jellyfish} \label{matmethods:jellyfish}
The breakthrough reconnections can also be encountered in biological systems. The gastrovascular system of the jellyfish \textit{Aurelia} is composed of canals in which seawater flows, carrying nutrients and oxygen to the surrounding tissues. New canals (sprouts) appear on the circulatory canal at the rim of the jellyfish sub-umbrella and grow toward four stomachs in the center of the jellyfish. We observed that the smaller sprouts reconnect to the long canal that just connected to the gastrocircular groove around the stomach~\cite{song2023morphogenesis} (Fig.~\ref{examples}e).

Jellyfish were reared at room temperature ($22^\circ$C) in artificial seawater prepared by diluting 28 g of synthetic sea salt (Instant Ocean; Spectrum Brands, Madison, WI) per liter of osmosis water (osmolarity 1100 mOsm). In the laboratory, polyps of the Roscoff strain~\cite{khalturin2019medusozoan} are used (obtained by courtesy of Konstantin Khalturin from the Marine Genomics Unit, Okinawa Institute of Science and Technology Graduate University, Onna, Okinawa, Japan). A more detailed description of jellyfish rearing can be found in Ref.~\cite{song2023morphogenesis}.

To observe the jellyfish gastrovascular system, the jellyfish are caught from the aquarium approximately 3 hours after feeding them with artemia. In this way, the gastrovascular canals are colored orange by the digested artemia. They are anesthetized with magnesium chloride dissolved in seawater. Then, they are placed in a Petri dish in shallow seawater with the subumbrella facing up. The images are taken by transillumination using a Leica macro zoom (MACROFLUO LEICA Z16 APO S/No: 5763648) and a Photron Fastcam SA3 camera. The images were stitched using Adobe Photoshop. The canals in the images were highlighted with gray color.

\subsubsection{Streamer discharge}\label{matmethods:streamers}
Streamer discharges are fundamental building blocks of sparks and ligthnings. An experimental photo presented in Fig.~\ref{examples}d was obtained by courtesy of Sander Nijdam~\cite{nijdam2020physics}. Here the mobilities are the conductivities inside the discharge channel and the air. Breakthrough reconnections in this context have been observed in many experiments~\cite{veldhuizen2002pulsed,winands2006temporal,briels2006circuit}. In Ref.~\cite{nijdam2009reconnection} stereo photography was used to fully reconstruct the three-dimensional structure of the reconnection events.

\subsubsection{Velocity of the fingers}
To estimate the velocity of the fingers in the experiments we binarized the images and prepared a mask that covered the area of the image in which the finger was growing. Next, we calculated the area of the shorter finger (within the mask) in successive frames. The approximate velocity of the finger was defined as the rate of change of this area. Finally, the velocity was smoothed using the Savitzky-Golay filter (function \textit{savgol\_filter} from the Scipy package).

\subsection{Numerical simulations}\label{matmethods:numerical}
The numerical calculations and simulations were performed using the codes available on the GitHub repository (\url{https://github.com/stzukowski/reticuler})~\cite{zukowski2022history, harris2020array, virtanen2020scipy, hunter2007matplotlib}. In the temporal simulations, the fingers have a constant width and end with semicircular tips, as in the static numerical calculations. In each time step of the evolution we solve the equations for the field~(Eqs.~(\ref{eq1})-(\ref{eq_last})) with the finite element method implemented in the FreeFEM++ software~\cite{hecht2012new}. To calculate the field in the system, we decompose it into two domains: the fingers and the outside. The mobility field is declared, $M(x,y)$, which is equal to $\lambda_1$ inside the fingers and $\lambda_2$ outside. Our script solves the equation for the field in the form: $\nabla \cdot (M(x,y) \nabla \phi (x,y))=0$.

To determine the velocity of the finger, we integrate the flux entering its tip. Following the principle of local symmetry~\cite{cohen2015path, zukowski2022history}, we extend the finger in the direction from which the highest flux comes (while keeping its width constant). The dynamical simulation in Fig.~\ref{velocity} was initiated with branches of length $l_\textrm{s}=0.04$ and $l_\textrm{l}=0.05$, while the simulations in Fig.~\ref{velocity_exps} started with branches of lengths: $l_\textrm{s}=0.002$ and $l_\textrm{l}=0.18$. The velocity in $y$ direction in Figs.~\ref{velocity} and \ref{velocity_exps} was treated with the Savitzky-Golay filter (\textit{savgol\_filter} function from the Scipy package) to smooth the numerical noise caused by the small order of magnitude of this velocity.

\begin{acknowledgments}
This project was partially supported by the National Science Center (Poland) under research grant 2020/02/Y/ST3/00121 and by the Mission for Transversal and Interdisciplinary Initiatives (MITI) of the French National Center of Scientific Research (CNRS) in the program Auto-Organisation 2021 and 2022 under research grants 212680, 284366, 238661. We also acknowledge the funding from the Polish National Agency for Academic Exchange and Campus France under grant BPN/BFR/2022/1/00033 in the program PHC Polonium. We also thank Sélène Jeammet for her help with the Saffman-Taylor experiment and Solène Song for the jellyfish timelapse images.
\end{acknowledgments}


\begin{thebibliography}{45}%
\makeatletter
\providecommand \@ifxundefined [1]{%
 \@ifx{#1\undefined}
}%
\providecommand \@ifnum [1]{%
 \ifnum #1\expandafter \@firstoftwo
 \else \expandafter \@secondoftwo
 \fi
}%
\providecommand \@ifx [1]{%
 \ifx #1\expandafter \@firstoftwo
 \else \expandafter \@secondoftwo
 \fi
}%
\providecommand \natexlab [1]{#1}%
\providecommand \enquote  [1]{``#1''}%
\providecommand \bibnamefont  [1]{#1}%
\providecommand \bibfnamefont [1]{#1}%
\providecommand \citenamefont [1]{#1}%
\providecommand \href@noop [0]{\@secondoftwo}%
\providecommand \href [0]{\begingroup \@sanitize@url \@href}%
\providecommand \@href[1]{\@@startlink{#1}\@@href}%
\providecommand \@@href[1]{\endgroup#1\@@endlink}%
\providecommand \@sanitize@url [0]{\catcode `\\12\catcode `\$12\catcode `\&12\catcode `\#12\catcode `\^12\catcode `\_12\catcode `\%12\relax}%
\providecommand \@@startlink[1]{}%
\providecommand \@@endlink[0]{}%
\providecommand \url  [0]{\begingroup\@sanitize@url \@url }%
\providecommand \@url [1]{\endgroup\@href {#1}{\urlprefix }}%
\providecommand \urlprefix  [0]{URL }%
\providecommand \Eprint [0]{\href }%
\providecommand \doibase [0]{https://doi.org/}%
\providecommand \selectlanguage [0]{\@gobble}%
\providecommand \bibinfo  [0]{\@secondoftwo}%
\providecommand \bibfield  [0]{\@secondoftwo}%
\providecommand \translation [1]{[#1]}%
\providecommand \BibitemOpen [0]{}%
\providecommand \bibitemStop [0]{}%
\providecommand \bibitemNoStop [0]{.\EOS\space}%
\providecommand \EOS [0]{\spacefactor3000\relax}%
\providecommand \BibitemShut  [1]{\csname bibitem#1\endcsname}%
\let\auto@bib@innerbib\@empty
\bibitem [{\citenamefont {Song}\ \emph {et~al.}(2023)\citenamefont {Song}, \citenamefont {{\.Z}ukowski}, \citenamefont {Gambini}, \citenamefont {Dantan}, \citenamefont {Mauroy}, \citenamefont {Douady},\ and\ \citenamefont {Cornelissen}}]{song2023morphogenesis}%
  \BibitemOpen
  \bibfield  {author} {\bibinfo {author} {\bibfnamefont {S.}~\bibnamefont {Song}}, \bibinfo {author} {\bibfnamefont {S.}~\bibnamefont {{\.Z}ukowski}}, \bibinfo {author} {\bibfnamefont {C.}~\bibnamefont {Gambini}}, \bibinfo {author} {\bibfnamefont {P.}~\bibnamefont {Dantan}}, \bibinfo {author} {\bibfnamefont {B.}~\bibnamefont {Mauroy}}, \bibinfo {author} {\bibfnamefont {S.}~\bibnamefont {Douady}},\ and\ \bibinfo {author} {\bibfnamefont {A.~J.~M.}\ \bibnamefont {Cornelissen}},\ }\bibfield  {title} {\bibinfo {title} {Morphogenesis of the gastrovascular canal network in {{Aurelia}} jellyfish: {{Variability}} and possible mechanisms},\ }\bibfield  {journal} {\bibinfo  {journal} {Frontiers in Physics}\ }\textbf {\bibinfo {volume} {10}},\ \href {https://doi.org/10.3389/fphy.2022.966327} {10.3389/fphy.2022.966327} (\bibinfo {year} {2023})\BibitemShut {NoStop}%
\bibitem [{\citenamefont {Mitchison}(1980)}]{mitchison1980model}%
  \BibitemOpen
  \bibfield  {author} {\bibinfo {author} {\bibfnamefont {G.~J.}\ \bibnamefont {Mitchison}},\ }\bibfield  {title} {\bibinfo {title} {A model for vein formation in higher plants},\ }\href@noop {} {\bibfield  {journal} {\bibinfo  {journal} {Proceedings of the Royal Society B}\ }\textbf {\bibinfo {volume} {207}},\ \bibinfo {pages} {79} (\bibinfo {year} {1980})}\BibitemShut {NoStop}%
\bibitem [{\citenamefont {Nguyen}\ \emph {et~al.}(2006)\citenamefont {Nguyen}, \citenamefont {Eichmann}, \citenamefont {Le~Noble},\ and\ \citenamefont {Fleury}}]{nguyen2006dynamics}%
  \BibitemOpen
  \bibfield  {author} {\bibinfo {author} {\bibfnamefont {T.-H.}\ \bibnamefont {Nguyen}}, \bibinfo {author} {\bibfnamefont {A.}~\bibnamefont {Eichmann}}, \bibinfo {author} {\bibfnamefont {F.}~\bibnamefont {Le~Noble}},\ and\ \bibinfo {author} {\bibfnamefont {V.}~\bibnamefont {Fleury}},\ }\bibfield  {title} {\bibinfo {title} {Dynamics of vascular branching morphogenesis: The effect of blood and tissue flow},\ }\href@noop {} {\bibfield  {journal} {\bibinfo  {journal} {Physical Review E}\ }\textbf {\bibinfo {volume} {73}},\ \bibinfo {pages} {061907} (\bibinfo {year} {2006})}\BibitemShut {NoStop}%
\bibitem [{\citenamefont {Petroff}\ \emph {et~al.}(2011)\citenamefont {Petroff}, \citenamefont {Devauchelle}, \citenamefont {Abrams}, \citenamefont {Lobkovsky}, \citenamefont {Kudrolli},\ and\ \citenamefont {Rothman}}]{petroff2011geometry}%
  \BibitemOpen
  \bibfield  {author} {\bibinfo {author} {\bibfnamefont {A.~P.}\ \bibnamefont {Petroff}}, \bibinfo {author} {\bibfnamefont {O.}~\bibnamefont {Devauchelle}}, \bibinfo {author} {\bibfnamefont {D.~M.}\ \bibnamefont {Abrams}}, \bibinfo {author} {\bibfnamefont {A.~E.}\ \bibnamefont {Lobkovsky}}, \bibinfo {author} {\bibfnamefont {A.}~\bibnamefont {Kudrolli}},\ and\ \bibinfo {author} {\bibfnamefont {D.~H.}\ \bibnamefont {Rothman}},\ }\bibfield  {title} {\bibinfo {title} {Geometry of valley growth},\ }\href@noop {} {\bibfield  {journal} {\bibinfo  {journal} {Journal of Fluid Mechanics}\ }\textbf {\bibinfo {volume} {673}},\ \bibinfo {pages} {245} (\bibinfo {year} {2011})}\BibitemShut {NoStop}%
\bibitem [{\citenamefont {Devauchelle}\ \emph {et~al.}(2012)\citenamefont {Devauchelle}, \citenamefont {Petroff}, \citenamefont {Seybold},\ and\ \citenamefont {Rothman}}]{devauchelle2012ramification}%
  \BibitemOpen
  \bibfield  {author} {\bibinfo {author} {\bibfnamefont {O.}~\bibnamefont {Devauchelle}}, \bibinfo {author} {\bibfnamefont {A.~P.}\ \bibnamefont {Petroff}}, \bibinfo {author} {\bibfnamefont {H.~J.}\ \bibnamefont {Seybold}},\ and\ \bibinfo {author} {\bibfnamefont {D.~H.}\ \bibnamefont {Rothman}},\ }\bibfield  {title} {\bibinfo {title} {Ramification of stream networks},\ }\href@noop {} {\bibfield  {journal} {\bibinfo  {journal} {Proceedings of the National Academy of Sciences}\ }\textbf {\bibinfo {volume} {109}},\ \bibinfo {pages} {20832} (\bibinfo {year} {2012})}\BibitemShut {NoStop}%
\bibitem [{\citenamefont {Konkol}\ \emph {et~al.}(2022)\citenamefont {Konkol}, \citenamefont {Schwenk}, \citenamefont {Katifori},\ and\ \citenamefont {Shaw}}]{konkol2022interplay}%
  \BibitemOpen
  \bibfield  {author} {\bibinfo {author} {\bibfnamefont {A.}~\bibnamefont {Konkol}}, \bibinfo {author} {\bibfnamefont {J.}~\bibnamefont {Schwenk}}, \bibinfo {author} {\bibfnamefont {E.}~\bibnamefont {Katifori}},\ and\ \bibinfo {author} {\bibfnamefont {J.~B.}\ \bibnamefont {Shaw}},\ }\bibfield  {title} {\bibinfo {title} {Interplay of {{River}} and {{Tidal Forcings Promotes Loops}} in {{Coastal Channel Networks}}},\ }\href {https://doi.org/10.1029/2022GL098284} {\bibfield  {journal} {\bibinfo  {journal} {Geophysical Research Letters}\ }\textbf {\bibinfo {volume} {49}},\ \bibinfo {pages} {e2022GL098284} (\bibinfo {year} {2022})}\BibitemShut {NoStop}%
\bibitem [{\citenamefont {Ke}\ \emph {et~al.}(2019)\citenamefont {Ke}, \citenamefont {Shaw}, \citenamefont {Mahon},\ and\ \citenamefont {Cathcart}}]{ke2019distributary}%
  \BibitemOpen
  \bibfield  {author} {\bibinfo {author} {\bibfnamefont {W.-T.}\ \bibnamefont {Ke}}, \bibinfo {author} {\bibfnamefont {J.~B.}\ \bibnamefont {Shaw}}, \bibinfo {author} {\bibfnamefont {R.~C.}\ \bibnamefont {Mahon}},\ and\ \bibinfo {author} {\bibfnamefont {C.~A.}\ \bibnamefont {Cathcart}},\ }\bibfield  {title} {\bibinfo {title} {Distributary {{Channel Networks}} as {{Moving Boundaries}}: {{Causes}} and {{Morphodynamic Effects}}},\ }\href {https://onlinelibrary.wiley.com/doi/abs/10.1029/2019JF005084} {\bibfield  {journal} {\bibinfo  {journal} {Journal of Geophysical Research}\ }\textbf {\bibinfo {volume} {124}},\ \bibinfo {pages} {1878} (\bibinfo {year} {2019})}\BibitemShut {NoStop}%
\bibitem [{\citenamefont {Szymczak}\ and\ \citenamefont {Ladd}(2011)}]{szymczak2011initial}%
  \BibitemOpen
  \bibfield  {author} {\bibinfo {author} {\bibfnamefont {P.}~\bibnamefont {Szymczak}}\ and\ \bibinfo {author} {\bibfnamefont {A.~J.~C.}\ \bibnamefont {Ladd}},\ }\bibfield  {title} {\bibinfo {title} {The initial stages of cave formation: {{Beyond}} the one-dimensional paradigm},\ }\href@noop {} {\bibfield  {journal} {\bibinfo  {journal} {Earth and Planetary Science Letters}\ }\textbf {\bibinfo {volume} {301}},\ \bibinfo {pages} {424} (\bibinfo {year} {2011})}\BibitemShut {NoStop}%
\bibitem [{\citenamefont {Ghani}\ \emph {et~al.}(2013)\citenamefont {Ghani}, \citenamefont {Koehn}, \citenamefont {Toussaint},\ and\ \citenamefont {Passchier}}]{ghani2013dynamic}%
  \BibitemOpen
  \bibfield  {author} {\bibinfo {author} {\bibfnamefont {I.}~\bibnamefont {Ghani}}, \bibinfo {author} {\bibfnamefont {D.}~\bibnamefont {Koehn}}, \bibinfo {author} {\bibfnamefont {R.}~\bibnamefont {Toussaint}},\ and\ \bibinfo {author} {\bibfnamefont {C.~W.}\ \bibnamefont {Passchier}},\ }\bibfield  {title} {\bibinfo {title} {Dynamic development of hydrofracture},\ }\href@noop {} {\bibfield  {journal} {\bibinfo  {journal} {Pure and Applied Geophysics}\ }\textbf {\bibinfo {volume} {170}},\ \bibinfo {pages} {1685} (\bibinfo {year} {2013})}\BibitemShut {NoStop}%
\bibitem [{\citenamefont {Katifori}\ \emph {et~al.}(2010)\citenamefont {Katifori}, \citenamefont {Sz{\"o}ll{\H o}si},\ and\ \citenamefont {Magnasco}}]{katifori2010damage}%
  \BibitemOpen
  \bibfield  {author} {\bibinfo {author} {\bibfnamefont {E.}~\bibnamefont {Katifori}}, \bibinfo {author} {\bibfnamefont {G.~J.}\ \bibnamefont {Sz{\"o}ll{\H o}si}},\ and\ \bibinfo {author} {\bibfnamefont {M.~O.}\ \bibnamefont {Magnasco}},\ }\bibfield  {title} {\bibinfo {title} {Damage and fluctuations induce loops in optimal transport networks},\ }\href@noop {} {\bibfield  {journal} {\bibinfo  {journal} {Physical Review Letters}\ }\textbf {\bibinfo {volume} {104}},\ \bibinfo {pages} {048704} (\bibinfo {year} {2010})}\BibitemShut {NoStop}%
\bibitem [{\citenamefont {Boyce}\ and\ \citenamefont {Knoll}(2002)}]{boyce2002evolution}%
  \BibitemOpen
  \bibfield  {author} {\bibinfo {author} {\bibfnamefont {C.~K.}\ \bibnamefont {Boyce}}\ and\ \bibinfo {author} {\bibfnamefont {A.~H.}\ \bibnamefont {Knoll}},\ }\bibfield  {title} {\bibinfo {title} {Evolution of developmental potential and the multiple independent origins of leaves in {{Paleozoic}} vascular plants},\ }\href@noop {} {\bibfield  {journal} {\bibinfo  {journal} {Paleobiology}\ ,\ \bibinfo {pages} {70}} (\bibinfo {year} {2002})}\BibitemShut {NoStop}%
\bibitem [{\citenamefont {Boyce}(2005)}]{boyce2005evolutionary}%
  \BibitemOpen
  \bibfield  {author} {\bibinfo {author} {\bibfnamefont {C.~K.}\ \bibnamefont {Boyce}},\ }\bibfield  {title} {\bibinfo {title} {The evolutionary history of roots and leaves},\ }in\ \href {https://doi.org/10.1016/B978-012088457-5/50025-3} {\emph {\bibinfo {booktitle} {Vascular Transport in Plants}}},\ \bibinfo {series and number} {Physiological Ecology},\ \bibinfo {editor} {edited by\ \bibinfo {editor} {\bibfnamefont {N.~M.}\ \bibnamefont {Holbrook}}\ and\ \bibinfo {editor} {\bibfnamefont {M.~A.}\ \bibnamefont {Zwieniecki}}}\ (\bibinfo  {publisher} {Academic Press},\ \bibinfo {address} {Burlington},\ \bibinfo {year} {2005})\ pp.\ \bibinfo {pages} {479--499}\BibitemShut {NoStop}%
\bibitem [{\citenamefont {Douady}\ \emph {et~al.}(2020)\citenamefont {Douady}, \citenamefont {Lagesse}, \citenamefont {Atashinbar}, \citenamefont {Bonnin}, \citenamefont {Pousse},\ and\ \citenamefont {Valcke}}]{douady2020work}%
  \BibitemOpen
  \bibfield  {author} {\bibinfo {author} {\bibfnamefont {S.}~\bibnamefont {Douady}}, \bibinfo {author} {\bibfnamefont {C.}~\bibnamefont {Lagesse}}, \bibinfo {author} {\bibfnamefont {M.}~\bibnamefont {Atashinbar}}, \bibinfo {author} {\bibfnamefont {P.}~\bibnamefont {Bonnin}}, \bibinfo {author} {\bibfnamefont {R.}~\bibnamefont {Pousse}},\ and\ \bibinfo {author} {\bibfnamefont {P.}~\bibnamefont {Valcke}},\ }\bibfield  {title} {\bibinfo {title} {A work on reticulated patterns},\ }\href {https://doi.org/10.5802 / crmeca.47} {\bibfield  {journal} {\bibinfo  {journal} {Comptes Rendus. M{\'e}canique}\ }\textbf {\bibinfo {volume} {348}},\ \bibinfo {pages} {659} (\bibinfo {year} {2020})}\BibitemShut {NoStop}%
\bibitem [{\citenamefont {Banavar}\ \emph {et~al.}(2000)\citenamefont {Banavar}, \citenamefont {Colaiori}, \citenamefont {Flammini}, \citenamefont {Maritan},\ and\ \citenamefont {Rinaldo}}]{banavar2000topology}%
  \BibitemOpen
  \bibfield  {author} {\bibinfo {author} {\bibfnamefont {J.~R.}\ \bibnamefont {Banavar}}, \bibinfo {author} {\bibfnamefont {F.}~\bibnamefont {Colaiori}}, \bibinfo {author} {\bibfnamefont {A.}~\bibnamefont {Flammini}}, \bibinfo {author} {\bibfnamefont {A.}~\bibnamefont {Maritan}},\ and\ \bibinfo {author} {\bibfnamefont {A.}~\bibnamefont {Rinaldo}},\ }\bibfield  {title} {\bibinfo {title} {Topology of the {{Fittest Transportation Network}}},\ }\href {https://doi.org/10.1103/PhysRevLett.84.4745} {\bibfield  {journal} {\bibinfo  {journal} {Physical Review Letters}\ }\textbf {\bibinfo {volume} {84}},\ \bibinfo {pages} {4745} (\bibinfo {year} {2000})}\BibitemShut {NoStop}%
\bibitem [{\citenamefont {Bohn}\ and\ \citenamefont {Magnasco}(2007)}]{bohn2007structure}%
  \BibitemOpen
  \bibfield  {author} {\bibinfo {author} {\bibfnamefont {S.}~\bibnamefont {Bohn}}\ and\ \bibinfo {author} {\bibfnamefont {M.~O.}\ \bibnamefont {Magnasco}},\ }\bibfield  {title} {\bibinfo {title} {Structure, {{Scaling}}, and {{Phase Transition}} in the {{Optimal Transport Network}}},\ }\href {https://doi.org/10.1103/PhysRevLett.98.088702} {\bibfield  {journal} {\bibinfo  {journal} {Physical Review Letters}\ }\textbf {\bibinfo {volume} {98}},\ \bibinfo {pages} {088702} (\bibinfo {year} {2007})}\BibitemShut {NoStop}%
\bibitem [{\citenamefont {Durand}(2007)}]{durand2007structure}%
  \BibitemOpen
  \bibfield  {author} {\bibinfo {author} {\bibfnamefont {M.}~\bibnamefont {Durand}},\ }\bibfield  {title} {\bibinfo {title} {Structure of {{Optimal Transport Networks Subject}} to a {{Global Constraint}}},\ }\href {https://doi.org/10.1103/PhysRevLett.98.088701} {\bibfield  {journal} {\bibinfo  {journal} {Physical Review Letters}\ }\textbf {\bibinfo {volume} {98}},\ \bibinfo {pages} {088701} (\bibinfo {year} {2007})}\BibitemShut {NoStop}%
\bibitem [{\citenamefont {Ronellenfitsch}\ and\ \citenamefont {Katifori}(2016)}]{ronellenfitsch2016global}%
  \BibitemOpen
  \bibfield  {author} {\bibinfo {author} {\bibfnamefont {H.}~\bibnamefont {Ronellenfitsch}}\ and\ \bibinfo {author} {\bibfnamefont {E.}~\bibnamefont {Katifori}},\ }\bibfield  {title} {\bibinfo {title} {Global optimization, local adaptation, and the role of growth in distribution networks},\ }\href@noop {} {\bibfield  {journal} {\bibinfo  {journal} {Physical Review Letters}\ }\textbf {\bibinfo {volume} {117}},\ \bibinfo {pages} {138301} (\bibinfo {year} {2016})}\BibitemShut {NoStop}%
\bibitem [{\citenamefont {Kaiser}\ \emph {et~al.}(2020)\citenamefont {Kaiser}, \citenamefont {Ronellenfitsch},\ and\ \citenamefont {Witthaut}}]{kaiser2020discontinuous}%
  \BibitemOpen
  \bibfield  {author} {\bibinfo {author} {\bibfnamefont {F.}~\bibnamefont {Kaiser}}, \bibinfo {author} {\bibfnamefont {H.}~\bibnamefont {Ronellenfitsch}},\ and\ \bibinfo {author} {\bibfnamefont {D.}~\bibnamefont {Witthaut}},\ }\bibfield  {title} {\bibinfo {title} {Discontinuous transition to loop formation in optimal supply networks},\ }\href {https://doi.org/10.1038/s41467-020-19567-2} {\bibfield  {journal} {\bibinfo  {journal} {Nature Communications}\ }\textbf {\bibinfo {volume} {11}},\ \bibinfo {pages} {5796} (\bibinfo {year} {2020})}\BibitemShut {NoStop}%
\bibitem [{\citenamefont {Couder}\ \emph {et~al.}(2002)\citenamefont {Couder}, \citenamefont {Pauchard}, \citenamefont {Allain}, \citenamefont {{Adda-Bedia}},\ and\ \citenamefont {Douady}}]{couder2002leaf}%
  \BibitemOpen
  \bibfield  {author} {\bibinfo {author} {\bibfnamefont {Y.}~\bibnamefont {Couder}}, \bibinfo {author} {\bibfnamefont {L.}~\bibnamefont {Pauchard}}, \bibinfo {author} {\bibfnamefont {C.}~\bibnamefont {Allain}}, \bibinfo {author} {\bibfnamefont {M.}~\bibnamefont {{Adda-Bedia}}},\ and\ \bibinfo {author} {\bibfnamefont {S.}~\bibnamefont {Douady}},\ }\bibfield  {title} {\bibinfo {title} {The leaf venation as formed in a tensorial field},\ }\href@noop {} {\bibfield  {journal} {\bibinfo  {journal} {The European Physical Journal B}\ }\textbf {\bibinfo {volume} {28}},\ \bibinfo {pages} {135} (\bibinfo {year} {2002})}\BibitemShut {NoStop}%
\bibitem [{\citenamefont {Bohn}\ \emph {et~al.}(2005)\citenamefont {Bohn}, \citenamefont {Pauchard},\ and\ \citenamefont {Couder}}]{bohn2005hierarchical}%
  \BibitemOpen
  \bibfield  {author} {\bibinfo {author} {\bibfnamefont {S.}~\bibnamefont {Bohn}}, \bibinfo {author} {\bibfnamefont {L.}~\bibnamefont {Pauchard}},\ and\ \bibinfo {author} {\bibfnamefont {Y.}~\bibnamefont {Couder}},\ }\bibfield  {title} {\bibinfo {title} {Hierarchical crack pattern as formed by successive domain divisions.},\ }\href {https://doi.org/10.1103/PhysRevE.71.046214} {\bibfield  {journal} {\bibinfo  {journal} {Physical Review E}\ }\textbf {\bibinfo {volume} {71}},\ \bibinfo {pages} {046214} (\bibinfo {year} {2005})}\BibitemShut {NoStop}%
\bibitem [{\citenamefont {Fleury}\ \emph {et~al.}(2001)\citenamefont {Fleury}, \citenamefont {Gouyet},\ and\ \citenamefont {Leonetti}}]{fleury2001branching}%
  \BibitemOpen
  \bibfield  {author} {\bibinfo {author} {\bibfnamefont {V.}~\bibnamefont {Fleury}}, \bibinfo {author} {\bibfnamefont {J.-F.}\ \bibnamefont {Gouyet}},\ and\ \bibinfo {author} {\bibfnamefont {M.}~\bibnamefont {Leonetti}},\ }\href@noop {} {\emph {\bibinfo {title} {Branching in Nature: Dynamics and Morphogenesis of Branching Structures, from Cell to River Networks}}},\ Vol.~\bibinfo {volume} {14}\ (\bibinfo  {publisher} {Springer Science \& Business Media},\ \bibinfo {year} {2001})\BibitemShut {NoStop}%
\bibitem [{\citenamefont {Ball}(2009)}]{ball2009branches}%
  \BibitemOpen
  \bibfield  {author} {\bibinfo {author} {\bibfnamefont {P.}~\bibnamefont {Ball}},\ }\href@noop {} {\emph {\bibinfo {title} {Branches: {{Nature}}'s Patterns: A Tapestry in Three Parts}}}\ (\bibinfo  {publisher} {OUP Oxford},\ \bibinfo {year} {2009})\BibitemShut {NoStop}%
\bibitem [{\citenamefont {Meakin}(1998)}]{meakin1998fractals}%
  \BibitemOpen
  \bibfield  {author} {\bibinfo {author} {\bibfnamefont {P.}~\bibnamefont {Meakin}},\ }\href@noop {} {\emph {\bibinfo {title} {Fractals, Scaling and Growth Far from Equilibrium}}},\ Vol.~\bibinfo {volume} {5}\ (\bibinfo  {publisher} {Cambridge University Press},\ \bibinfo {year} {1998})\BibitemShut {NoStop}%
\bibitem [{\citenamefont {Detwiler}\ \emph {et~al.}(2003)\citenamefont {Detwiler}, \citenamefont {Glass},\ and\ \citenamefont {Bourcier}}]{detwiler2003experimental}%
  \BibitemOpen
  \bibfield  {author} {\bibinfo {author} {\bibfnamefont {R.~L.}\ \bibnamefont {Detwiler}}, \bibinfo {author} {\bibfnamefont {R.~J.}\ \bibnamefont {Glass}},\ and\ \bibinfo {author} {\bibfnamefont {W.~L.}\ \bibnamefont {Bourcier}},\ }\bibfield  {title} {\bibinfo {title} {Experimental observations of fracture dissolution: {{The}} role of {{Peclet}} number on evolving aperture variability},\ }\href@noop {} {\bibfield  {journal} {\bibinfo  {journal} {Geophysical Research Letters}\ }\textbf {\bibinfo {volume} {30}} (\bibinfo {year} {2003})}\BibitemShut {NoStop}%
\bibitem [{\citenamefont {Starchenko}\ \emph {et~al.}(2016)\citenamefont {Starchenko}, \citenamefont {Marra},\ and\ \citenamefont {Ladd}}]{starchenko2016three}%
  \BibitemOpen
  \bibfield  {author} {\bibinfo {author} {\bibfnamefont {V.}~\bibnamefont {Starchenko}}, \bibinfo {author} {\bibfnamefont {C.~J.}\ \bibnamefont {Marra}},\ and\ \bibinfo {author} {\bibfnamefont {A.~J.~C.}\ \bibnamefont {Ladd}},\ }\bibfield  {title} {\bibinfo {title} {Three-dimensional simulations of fracture dissolution},\ }\href@noop {} {\bibfield  {journal} {\bibinfo  {journal} {Journal of Geophysical Research}\ }\textbf {\bibinfo {volume} {121}},\ \bibinfo {pages} {6421} (\bibinfo {year} {2016})}\BibitemShut {NoStop}%
\bibitem [{\citenamefont {Yang}\ \emph {et~al.}(2022)\citenamefont {Yang}, \citenamefont {Kong}, \citenamefont {Cao}, \citenamefont {Wu},\ and\ \citenamefont {Li}}]{yang2022hydrodynamics}%
  \BibitemOpen
  \bibfield  {author} {\bibinfo {author} {\bibfnamefont {S.}~\bibnamefont {Yang}}, \bibinfo {author} {\bibfnamefont {G.}~\bibnamefont {Kong}}, \bibinfo {author} {\bibfnamefont {Z.}~\bibnamefont {Cao}}, \bibinfo {author} {\bibfnamefont {Z.}~\bibnamefont {Wu}},\ and\ \bibinfo {author} {\bibfnamefont {H.}~\bibnamefont {Li}},\ }\bibfield  {title} {\bibinfo {title} {Hydrodynamics of gas-liquid displacement in porous media: Fingering pattern evolution at the breakthrough moment and the stable state},\ }\href {https://doi.org/10.1016/j.advwatres.2022.104324} {\bibfield  {journal} {\bibinfo  {journal} {Advances in Water Resources}\ ,\ \bibinfo {pages} {104324}} (\bibinfo {year} {2022})}\BibitemShut {NoStop}%
\bibitem [{\citenamefont {Nijdam}\ \emph {et~al.}(2009)\citenamefont {Nijdam}, \citenamefont {Geurts}, \citenamefont {{van Veldhuizen}},\ and\ \citenamefont {Ebert}}]{nijdam2009reconnection}%
  \BibitemOpen
  \bibfield  {author} {\bibinfo {author} {\bibfnamefont {S.}~\bibnamefont {Nijdam}}, \bibinfo {author} {\bibfnamefont {C.~G.~C.}\ \bibnamefont {Geurts}}, \bibinfo {author} {\bibfnamefont {E.~M.}\ \bibnamefont {{van Veldhuizen}}},\ and\ \bibinfo {author} {\bibfnamefont {U.}~\bibnamefont {Ebert}},\ }\bibfield  {title} {\bibinfo {title} {Reconnection and merging of positive streamers in air},\ }\href {https://doi.org/10.1088/0022-3727/42/4/045201} {\bibfield  {journal} {\bibinfo  {journal} {Journal of Physics D}\ }\textbf {\bibinfo {volume} {42}},\ \bibinfo {pages} {045201} (\bibinfo {year} {2009})}\BibitemShut {NoStop}%
\bibitem [{\citenamefont {Muskat}(1937)}]{muskat1937flow}%
  \BibitemOpen
  \bibfield  {author} {\bibinfo {author} {\bibfnamefont {M.}~\bibnamefont {Muskat}},\ }\bibfield  {title} {\bibinfo {title} {The flow of fluids through porous media},\ }\href@noop {} {\bibfield  {journal} {\bibinfo  {journal} {Journal of Applied Physics}\ }\textbf {\bibinfo {volume} {8}},\ \bibinfo {pages} {274} (\bibinfo {year} {1937})}\BibitemShut {NoStop}%
\bibitem [{\citenamefont {Stevens}(1974)}]{stevens1974patterns}%
  \BibitemOpen
  \bibfield  {author} {\bibinfo {author} {\bibfnamefont {P.~S.}\ \bibnamefont {Stevens}},\ }\href@noop {} {\emph {\bibinfo {title} {Patterns in {{Nature}}}}},\ \bibinfo {edition} {1st}\ ed.\ (\bibinfo  {publisher} {{Little, Brown and Company}},\ \bibinfo {year} {1974})\BibitemShut {NoStop}%
\bibitem [{\citenamefont {Luque}\ and\ \citenamefont {Ebert}(2014)}]{luque2014growing}%
  \BibitemOpen
  \bibfield  {author} {\bibinfo {author} {\bibfnamefont {A.}~\bibnamefont {Luque}}\ and\ \bibinfo {author} {\bibfnamefont {U.}~\bibnamefont {Ebert}},\ }\bibfield  {title} {\bibinfo {title} {Growing discharge trees with self-consistent charge transport: The collective dynamics of streamers},\ }\href@noop {} {\bibfield  {journal} {\bibinfo  {journal} {New Journal of Physics}\ }\textbf {\bibinfo {volume} {16}},\ \bibinfo {pages} {013039} (\bibinfo {year} {2014})}\BibitemShut {NoStop}%
\bibitem [{\citenamefont {Budek}\ \emph {et~al.}(2017)\citenamefont {Budek}, \citenamefont {Kwiatkowski},\ and\ \citenamefont {Szymczak}}]{budek2017effect}%
  \BibitemOpen
  \bibfield  {author} {\bibinfo {author} {\bibfnamefont {A.}~\bibnamefont {Budek}}, \bibinfo {author} {\bibfnamefont {K.}~\bibnamefont {Kwiatkowski}},\ and\ \bibinfo {author} {\bibfnamefont {P.}~\bibnamefont {Szymczak}},\ }\bibfield  {title} {\bibinfo {title} {Effect of mobility ratio on interaction between the fingers in unstable growth processes},\ }\href@noop {} {\bibfield  {journal} {\bibinfo  {journal} {Physical Review E}\ }\textbf {\bibinfo {volume} {96}},\ \bibinfo {pages} {042218} (\bibinfo {year} {2017})}\BibitemShut {NoStop}%
\bibitem [{\citenamefont {Hecht}(2012)}]{hecht2012new}%
  \BibitemOpen
  \bibfield  {author} {\bibinfo {author} {\bibfnamefont {F.}~\bibnamefont {Hecht}},\ }\bibfield  {title} {\bibinfo {title} {New development in {{FreeFem}}++},\ }\href {https://freefem.org/} {\bibfield  {journal} {\bibinfo  {journal} {Journal of Numerical Mathematics}\ }\textbf {\bibinfo {volume} {20}},\ \bibinfo {pages} {251} (\bibinfo {year} {2012})}\BibitemShut {NoStop}%
\bibitem [{\citenamefont {Bischofberger}\ \emph {et~al.}(2015)\citenamefont {Bischofberger}, \citenamefont {Ramachandran},\ and\ \citenamefont {Nagel}}]{bischofberger2015island}%
  \BibitemOpen
  \bibfield  {author} {\bibinfo {author} {\bibfnamefont {I.}~\bibnamefont {Bischofberger}}, \bibinfo {author} {\bibfnamefont {R.}~\bibnamefont {Ramachandran}},\ and\ \bibinfo {author} {\bibfnamefont {S.~R.}\ \bibnamefont {Nagel}},\ }\bibfield  {title} {\bibinfo {title} {An island of stability in a sea of fingers: Emergent global features of the viscous-flow instability},\ }\href@noop {} {\bibfield  {journal} {\bibinfo  {journal} {Soft Matter}\ }\textbf {\bibinfo {volume} {11}},\ \bibinfo {pages} {7428} (\bibinfo {year} {2015})}\BibitemShut {NoStop}%
\bibitem [{\citenamefont {Zik}\ and\ \citenamefont {Moses}(1999)}]{zik1999fingering}%
  \BibitemOpen
  \bibfield  {author} {\bibinfo {author} {\bibfnamefont {O.}~\bibnamefont {Zik}}\ and\ \bibinfo {author} {\bibfnamefont {E.}~\bibnamefont {Moses}},\ }\bibfield  {title} {\bibinfo {title} {Fingering instability in combustion: An extended view},\ }\href@noop {} {\bibfield  {journal} {\bibinfo  {journal} {Physical Review E}\ }\textbf {\bibinfo {volume} {60}},\ \bibinfo {pages} {518} (\bibinfo {year} {1999})}\BibitemShut {NoStop}%
\bibitem [{\citenamefont {Osselin}\ \emph {et~al.}(2016)\citenamefont {Osselin}, \citenamefont {Kondratiuk}, \citenamefont {Budek}, \citenamefont {Cybulski}, \citenamefont {Garstecki},\ and\ \citenamefont {Szymczak}}]{osselin2016microfluidic}%
  \BibitemOpen
  \bibfield  {author} {\bibinfo {author} {\bibfnamefont {F.}~\bibnamefont {Osselin}}, \bibinfo {author} {\bibfnamefont {P.}~\bibnamefont {Kondratiuk}}, \bibinfo {author} {\bibfnamefont {A.}~\bibnamefont {Budek}}, \bibinfo {author} {\bibfnamefont {O.}~\bibnamefont {Cybulski}}, \bibinfo {author} {\bibfnamefont {P.}~\bibnamefont {Garstecki}},\ and\ \bibinfo {author} {\bibfnamefont {P.}~\bibnamefont {Szymczak}},\ }\bibfield  {title} {\bibinfo {title} {Microfluidic observation of the onset of reactive-infitration instability in an analog fracture},\ }\href {https://doi.org/10.1002/2016GL069261} {\bibfield  {journal} {\bibinfo  {journal} {Geophysical Research Letters}\ }\textbf {\bibinfo {volume} {43}},\ \bibinfo {pages} {6907} (\bibinfo {year} {2016})}\BibitemShut {NoStop}%
\bibitem [{\citenamefont {Khalturin}\ \emph {et~al.}(2019)\citenamefont {Khalturin}, \citenamefont {Shinzato}, \citenamefont {Khalturina}, \citenamefont {Hamada}, \citenamefont {Fujie}, \citenamefont {Koyanagi}, \citenamefont {Kanda}, \citenamefont {Goto}, \citenamefont {{Anton-Erxleben}}, \citenamefont {Toyokawa}, \citenamefont {Toshino},\ and\ \citenamefont {Satoh}}]{khalturin2019medusozoan}%
  \BibitemOpen
  \bibfield  {author} {\bibinfo {author} {\bibfnamefont {K.}~\bibnamefont {Khalturin}}, \bibinfo {author} {\bibfnamefont {C.}~\bibnamefont {Shinzato}}, \bibinfo {author} {\bibfnamefont {M.}~\bibnamefont {Khalturina}}, \bibinfo {author} {\bibfnamefont {M.}~\bibnamefont {Hamada}}, \bibinfo {author} {\bibfnamefont {M.}~\bibnamefont {Fujie}}, \bibinfo {author} {\bibfnamefont {R.}~\bibnamefont {Koyanagi}}, \bibinfo {author} {\bibfnamefont {M.}~\bibnamefont {Kanda}}, \bibinfo {author} {\bibfnamefont {H.}~\bibnamefont {Goto}}, \bibinfo {author} {\bibfnamefont {F.}~\bibnamefont {{Anton-Erxleben}}}, \bibinfo {author} {\bibfnamefont {M.}~\bibnamefont {Toyokawa}}, \bibinfo {author} {\bibfnamefont {S.}~\bibnamefont {Toshino}},\ and\ \bibinfo {author} {\bibfnamefont {N.}~\bibnamefont {Satoh}},\ }\bibfield  {title} {\bibinfo {title} {Medusozoan genomes inform the evolution of the jellyfish body plan},\ }\href {https://doi.org/10.1038/s41559-019-0853-y} {\bibfield  {journal} {\bibinfo  {journal} {Nature Ecology \&
  Evolution}\ }\textbf {\bibinfo {volume} {3}},\ \bibinfo {pages} {811} (\bibinfo {year} {2019})}\BibitemShut {NoStop}%
\bibitem [{\citenamefont {Nijdam}\ \emph {et~al.}(2020)\citenamefont {Nijdam}, \citenamefont {Teunissen},\ and\ \citenamefont {Ebert}}]{nijdam2020physics}%
  \BibitemOpen
  \bibfield  {author} {\bibinfo {author} {\bibfnamefont {S.}~\bibnamefont {Nijdam}}, \bibinfo {author} {\bibfnamefont {J.}~\bibnamefont {Teunissen}},\ and\ \bibinfo {author} {\bibfnamefont {U.}~\bibnamefont {Ebert}},\ }\bibfield  {title} {\bibinfo {title} {The physics of streamer discharge phenomena},\ }\href {https://doi.org/10.1088/1361-6595/abaa05} {\bibfield  {journal} {\bibinfo  {journal} {Plasma Sources Science and Technology}\ }\textbf {\bibinfo {volume} {29}},\ \bibinfo {pages} {103001} (\bibinfo {year} {2020})}\BibitemShut {NoStop}%
\bibitem [{\citenamefont {van Veldhuizen}\ and\ \citenamefont {Rutgers}(2002)}]{veldhuizen2002pulsed}%
  \BibitemOpen
  \bibfield  {author} {\bibinfo {author} {\bibfnamefont {E.~M.}\ \bibnamefont {van Veldhuizen}}\ and\ \bibinfo {author} {\bibfnamefont {W.~R.}\ \bibnamefont {Rutgers}},\ }\bibfield  {title} {\bibinfo {title} {Pulsed positive corona streamer propagation and branching},\ }\href {https://doi.org/10.1088/0022-3727/35/17/313} {\bibfield  {journal} {\bibinfo  {journal} {Journal of Physics D: Applied Physics}\ }\textbf {\bibinfo {volume} {35}},\ \bibinfo {pages} {2169} (\bibinfo {year} {2002})}\BibitemShut {NoStop}%
\bibitem [{\citenamefont {Winands}\ \emph {et~al.}(2006)\citenamefont {Winands}, \citenamefont {Liu}, \citenamefont {Pemen}, \citenamefont {van Heesch}, \citenamefont {Yan},\ and\ \citenamefont {van Veldhuizen}}]{winands2006temporal}%
  \BibitemOpen
  \bibfield  {author} {\bibinfo {author} {\bibfnamefont {G.~J.~J.}\ \bibnamefont {Winands}}, \bibinfo {author} {\bibfnamefont {Z.}~\bibnamefont {Liu}}, \bibinfo {author} {\bibfnamefont {A.~J.~M.}\ \bibnamefont {Pemen}}, \bibinfo {author} {\bibfnamefont {E.~J.~M.}\ \bibnamefont {van Heesch}}, \bibinfo {author} {\bibfnamefont {K.}~\bibnamefont {Yan}},\ and\ \bibinfo {author} {\bibfnamefont {E.~M.}\ \bibnamefont {van Veldhuizen}},\ }\bibfield  {title} {\bibinfo {title} {Temporal development and chemical efficiency of positive streamers in a large scale wire-plate reactor as a function of voltage waveform parameters},\ }\href {https://doi.org/10.1088/0022-3727/39/14/020} {\bibfield  {journal} {\bibinfo  {journal} {Journal of Physics D: Applied Physics}\ }\textbf {\bibinfo {volume} {39}},\ \bibinfo {pages} {3010} (\bibinfo {year} {2006})}\BibitemShut {NoStop}%
\bibitem [{\citenamefont {Briels}\ \emph {et~al.}(2006)\citenamefont {Briels}, \citenamefont {Kos}, \citenamefont {van Veldhuizen},\ and\ \citenamefont {Ebert}}]{briels2006circuit}%
  \BibitemOpen
  \bibfield  {author} {\bibinfo {author} {\bibfnamefont {T.~M.~P.}\ \bibnamefont {Briels}}, \bibinfo {author} {\bibfnamefont {J.}~\bibnamefont {Kos}}, \bibinfo {author} {\bibfnamefont {E.~M.}\ \bibnamefont {van Veldhuizen}},\ and\ \bibinfo {author} {\bibfnamefont {U.}~\bibnamefont {Ebert}},\ }\bibfield  {title} {\bibinfo {title} {Circuit dependence of the diameter of pulsed positive streamers in air},\ }\href {https://doi.org/10.1088/0022-3727/39/24/016} {\bibfield  {journal} {\bibinfo  {journal} {Journal of Physics D: Applied Physics}\ }\textbf {\bibinfo {volume} {39}},\ \bibinfo {pages} {5201} (\bibinfo {year} {2006})}\BibitemShut {NoStop}%
\bibitem [{\citenamefont {{\.Z}ukowski}\ \emph {et~al.}(2022)\citenamefont {{\.Z}ukowski}, \citenamefont {Morawiecki}, \citenamefont {Seybold},\ and\ \citenamefont {Szymczak}}]{zukowski2022history}%
  \BibitemOpen
  \bibfield  {author} {\bibinfo {author} {\bibfnamefont {S.}~\bibnamefont {{\.Z}ukowski}}, \bibinfo {author} {\bibfnamefont {P.}~\bibnamefont {Morawiecki}}, \bibinfo {author} {\bibfnamefont {H.~J.}\ \bibnamefont {Seybold}},\ and\ \bibinfo {author} {\bibfnamefont {P.}~\bibnamefont {Szymczak}},\ }\bibfield  {title} {\bibinfo {title} {Through history to growth dynamics: Deciphering the evolution of spatial networks},\ }\href {https://doi.org/10.1038/s41598-022-24656-x} {\bibfield  {journal} {\bibinfo  {journal} {Scientific Reports}\ }\textbf {\bibinfo {volume} {12}},\ \bibinfo {pages} {20407} (\bibinfo {year} {2022})}\BibitemShut {NoStop}%
\bibitem [{\citenamefont {Harris}\ \emph {et~al.}(2020)\citenamefont {Harris}, \citenamefont {Millman}, \citenamefont {{van der Walt}}, \citenamefont {Gommers}, \citenamefont {Virtanen}, \citenamefont {Cournapeau}, \citenamefont {Wieser}, \citenamefont {Taylor}, \citenamefont {Berg}, \citenamefont {Smith}, \citenamefont {Kern}, \citenamefont {Picus}, \citenamefont {Hoyer}, \citenamefont {{van Kerkwijk}}, \citenamefont {Brett}, \citenamefont {Haldane}, \citenamefont {{del R{\'i}o}}, \citenamefont {Wiebe}, \citenamefont {Peterson}, \citenamefont {{G{\'e}rard-Marchant}}, \citenamefont {Sheppard}, \citenamefont {Reddy}, \citenamefont {Weckesser}, \citenamefont {Abbasi}, \citenamefont {Gohlke},\ and\ \citenamefont {Oliphant}}]{harris2020array}%
  \BibitemOpen
  \bibfield  {author} {\bibinfo {author} {\bibfnamefont {C.~R.}\ \bibnamefont {Harris}}, \bibinfo {author} {\bibfnamefont {K.~J.}\ \bibnamefont {Millman}}, \bibinfo {author} {\bibfnamefont {S.~J.}\ \bibnamefont {{van der Walt}}}, \bibinfo {author} {\bibfnamefont {R.}~\bibnamefont {Gommers}}, \bibinfo {author} {\bibfnamefont {P.}~\bibnamefont {Virtanen}}, \bibinfo {author} {\bibfnamefont {D.}~\bibnamefont {Cournapeau}}, \bibinfo {author} {\bibfnamefont {E.}~\bibnamefont {Wieser}}, \bibinfo {author} {\bibfnamefont {J.}~\bibnamefont {Taylor}}, \bibinfo {author} {\bibfnamefont {S.}~\bibnamefont {Berg}}, \bibinfo {author} {\bibfnamefont {N.~J.}\ \bibnamefont {Smith}}, \bibinfo {author} {\bibfnamefont {R.}~\bibnamefont {Kern}}, \bibinfo {author} {\bibfnamefont {M.}~\bibnamefont {Picus}}, \bibinfo {author} {\bibfnamefont {S.}~\bibnamefont {Hoyer}}, \bibinfo {author} {\bibfnamefont {M.~H.}\ \bibnamefont {{van Kerkwijk}}}, \bibinfo {author} {\bibfnamefont {M.}~\bibnamefont {Brett}}, \bibinfo {author} {\bibfnamefont
  {A.}~\bibnamefont {Haldane}}, \bibinfo {author} {\bibfnamefont {J.~F.}\ \bibnamefont {{del R{\'i}o}}}, \bibinfo {author} {\bibfnamefont {M.}~\bibnamefont {Wiebe}}, \bibinfo {author} {\bibfnamefont {P.}~\bibnamefont {Peterson}}, \bibinfo {author} {\bibfnamefont {P.}~\bibnamefont {{G{\'e}rard-Marchant}}}, \bibinfo {author} {\bibfnamefont {K.}~\bibnamefont {Sheppard}}, \bibinfo {author} {\bibfnamefont {T.}~\bibnamefont {Reddy}}, \bibinfo {author} {\bibfnamefont {W.}~\bibnamefont {Weckesser}}, \bibinfo {author} {\bibfnamefont {H.}~\bibnamefont {Abbasi}}, \bibinfo {author} {\bibfnamefont {C.}~\bibnamefont {Gohlke}},\ and\ \bibinfo {author} {\bibfnamefont {T.~E.}\ \bibnamefont {Oliphant}},\ }\bibfield  {title} {\bibinfo {title} {Array programming with {{NumPy}}},\ }\href {https://doi.org/10.1038/s41586-020-2649-2} {\bibfield  {journal} {\bibinfo  {journal} {Nature}\ }\textbf {\bibinfo {volume} {585}},\ \bibinfo {pages} {357} (\bibinfo {year} {2020})}\BibitemShut {NoStop}%
\bibitem [{\citenamefont {Virtanen}\ \emph {et~al.}(2020)\citenamefont {Virtanen}, \citenamefont {Gommers}, \citenamefont {Oliphant}, \citenamefont {Haberland}, \citenamefont {Reddy}, \citenamefont {Cournapeau}, \citenamefont {Burovski}, \citenamefont {Peterson}, \citenamefont {Weckesser}, \citenamefont {Bright}, \citenamefont {{van der Walt}}, \citenamefont {Brett}, \citenamefont {Wilson}, \citenamefont {Millman}, \citenamefont {Mayorov}, \citenamefont {Nelson}, \citenamefont {Jones}, \citenamefont {Kern}, \citenamefont {Larson}, \citenamefont {Carey}, \citenamefont {Polat}, \citenamefont {Feng}, \citenamefont {Moore}, \citenamefont {VanderPlas}, \citenamefont {Laxalde}, \citenamefont {Perktold}, \citenamefont {Cimrman}, \citenamefont {Henriksen}, \citenamefont {Quintero}, \citenamefont {Harris}, \citenamefont {Archibald}, \citenamefont {Ribeiro}, \citenamefont {Pedregosa}, \citenamefont {{van Mulbregt}},\ and\ \citenamefont {{SciPy 1.0 Contributors}}}]{virtanen2020scipy}%
  \BibitemOpen
  \bibfield  {author} {\bibinfo {author} {\bibfnamefont {P.}~\bibnamefont {Virtanen}}, \bibinfo {author} {\bibfnamefont {R.}~\bibnamefont {Gommers}}, \bibinfo {author} {\bibfnamefont {T.~E.}\ \bibnamefont {Oliphant}}, \bibinfo {author} {\bibfnamefont {M.}~\bibnamefont {Haberland}}, \bibinfo {author} {\bibfnamefont {T.}~\bibnamefont {Reddy}}, \bibinfo {author} {\bibfnamefont {D.}~\bibnamefont {Cournapeau}}, \bibinfo {author} {\bibfnamefont {E.}~\bibnamefont {Burovski}}, \bibinfo {author} {\bibfnamefont {P.}~\bibnamefont {Peterson}}, \bibinfo {author} {\bibfnamefont {W.}~\bibnamefont {Weckesser}}, \bibinfo {author} {\bibfnamefont {J.}~\bibnamefont {Bright}}, \bibinfo {author} {\bibfnamefont {S.~J.}\ \bibnamefont {{van der Walt}}}, \bibinfo {author} {\bibfnamefont {M.}~\bibnamefont {Brett}}, \bibinfo {author} {\bibfnamefont {J.}~\bibnamefont {Wilson}}, \bibinfo {author} {\bibfnamefont {K.~J.}\ \bibnamefont {Millman}}, \bibinfo {author} {\bibfnamefont {N.}~\bibnamefont {Mayorov}}, \bibinfo {author} {\bibfnamefont
  {A.~R.~J.}\ \bibnamefont {Nelson}}, \bibinfo {author} {\bibfnamefont {E.}~\bibnamefont {Jones}}, \bibinfo {author} {\bibfnamefont {R.}~\bibnamefont {Kern}}, \bibinfo {author} {\bibfnamefont {E.}~\bibnamefont {Larson}}, \bibinfo {author} {\bibfnamefont {C.~J.}\ \bibnamefont {Carey}}, \bibinfo {author} {\bibfnamefont {{\.I}.}~\bibnamefont {Polat}}, \bibinfo {author} {\bibfnamefont {Y.}~\bibnamefont {Feng}}, \bibinfo {author} {\bibfnamefont {E.~W.}\ \bibnamefont {Moore}}, \bibinfo {author} {\bibfnamefont {J.}~\bibnamefont {VanderPlas}}, \bibinfo {author} {\bibfnamefont {D.}~\bibnamefont {Laxalde}}, \bibinfo {author} {\bibfnamefont {J.}~\bibnamefont {Perktold}}, \bibinfo {author} {\bibfnamefont {R.}~\bibnamefont {Cimrman}}, \bibinfo {author} {\bibfnamefont {I.}~\bibnamefont {Henriksen}}, \bibinfo {author} {\bibfnamefont {E.~A.}\ \bibnamefont {Quintero}}, \bibinfo {author} {\bibfnamefont {C.~R.}\ \bibnamefont {Harris}}, \bibinfo {author} {\bibfnamefont {A.~M.}\ \bibnamefont {Archibald}}, \bibinfo {author}
  {\bibfnamefont {A.~H.}\ \bibnamefont {Ribeiro}}, \bibinfo {author} {\bibfnamefont {F.}~\bibnamefont {Pedregosa}}, \bibinfo {author} {\bibfnamefont {P.}~\bibnamefont {{van Mulbregt}}},\ and\ \bibinfo {author} {\bibnamefont {{SciPy 1.0 Contributors}}},\ }\bibfield  {title} {\bibinfo {title} {{{SciPy}} 1.0: {{Fundamental}} algorithms for scientific computing in python},\ }\href {https://doi.org/10.1038/s41592-019-0686-2} {\bibfield  {journal} {\bibinfo  {journal} {Nature Methods}\ }\textbf {\bibinfo {volume} {17}},\ \bibinfo {pages} {261} (\bibinfo {year} {2020})}\BibitemShut {NoStop}%
\bibitem [{\citenamefont {Hunter}(2007)}]{hunter2007matplotlib}%
  \BibitemOpen
  \bibfield  {author} {\bibinfo {author} {\bibfnamefont {J.~D.}\ \bibnamefont {Hunter}},\ }\bibfield  {title} {\bibinfo {title} {Matplotlib: {{A 2D}} graphics environment},\ }\href {https://doi.org/10.1109/MCSE.2007.55} {\bibfield  {journal} {\bibinfo  {journal} {Computing in Science \& Engineering}\ }\textbf {\bibinfo {volume} {9}},\ \bibinfo {pages} {90} (\bibinfo {year} {2007})}\BibitemShut {NoStop}%
\bibitem [{\citenamefont {Cohen}\ \emph {et~al.}(2015)\citenamefont {Cohen}, \citenamefont {Devauchelle}, \citenamefont {Seybold}, \citenamefont {Yi}, \citenamefont {Szymczak},\ and\ \citenamefont {Rothman}}]{cohen2015path}%
  \BibitemOpen
  \bibfield  {author} {\bibinfo {author} {\bibfnamefont {Y.}~\bibnamefont {Cohen}}, \bibinfo {author} {\bibfnamefont {O.}~\bibnamefont {Devauchelle}}, \bibinfo {author} {\bibfnamefont {H.~J.}\ \bibnamefont {Seybold}}, \bibinfo {author} {\bibfnamefont {R.~S.}\ \bibnamefont {Yi}}, \bibinfo {author} {\bibfnamefont {P.}~\bibnamefont {Szymczak}},\ and\ \bibinfo {author} {\bibfnamefont {D.~H.}\ \bibnamefont {Rothman}},\ }\bibfield  {title} {\bibinfo {title} {Path selection in the growth of rivers},\ }\href@noop {} {\bibfield  {journal} {\bibinfo  {journal} {Proceedings of the National Academy of Sciences}\ }\textbf {\bibinfo {volume} {112}},\ \bibinfo {pages} {14132} (\bibinfo {year} {2015})}\BibitemShut {NoStop}%
\end{thebibliography}
%

\end{document}